\UseRawInputEncoding
\documentclass[a4paper,fleqn,usenatbib]{mnras}

\usepackage{subfigure}
\usepackage{epsfig}
\usepackage{rotating}
\usepackage[T1]{fontenc}
\usepackage{ae,aecompl}
\usepackage{tikz}

\usepackage{graphicx}	
\usepackage{amsmath}	
\usepackage{amssymb}	

\usepackage{color}
\usepackage{natbib}

\usepackage{booktabs}

\newcommand{\Ms}{{\ensuremath{\mathrm{M}_{\odot} }}}

\title[Collapse of Atomically Cooled Haloes I]{The collapse of atomically-cooled primordial haloes. I. High Lyman-Werner backgrounds.}

\author[Patrick et al.]{Samuel J. Patrick$^{1}$\thanks{E-mail: spatrick@ed.ac.uk}, 
Daniel J. Whalen$^{2}$, 
Muhammed A. Latif$^3$, 
Jacob S. Elford$^4$
\\
\\
$^{1}$Institute for Astronomy, University of Edinburgh, Royal Observatory, Blackford Hill, Edinburgh EH9 3HJ, UK \\
$^{2}$Institute of Cosmology and Gravitation, University of Portsmouth, Portsmouth PO1 
3FX, UK \\ 
$^3$Physics Department, College of Science, United Arab Emirates University, PO Box 15551, Al-Ain, UAE \\
$^4$School of Physics and Astronomy, Queen's Buildings, The Parade, Cardiff University, Cardiff, CF24 3AA, UK 
}

\date{Accepted XXX. Received YYY; in original form ZZZ}

\pubyear{2020}

\begin{document}
\label{firstpage}
\pagerange{\pageref{firstpage}--\pageref{lastpage}}
\maketitle

\begin{abstract}

Pristine, atomically-cooled haloes may be the sites of primordial quasar formation because atomic cooling triggers rapid baryon collapse that can create 10$^4$ - 10$^5$ \Ms\ black hole seeds. However, no numerical simulation has ever followed the collapse of these haloes for the times required to form supermassive stars and direct-collapse black holes (DCBHs). We have now modeled baryon collapse in atomically-cooled haloes with a wide range of spin parameters and assembly histories for times that are sufficient for DCBH formation. Fragmentation of accretion disks after $\sim$ 500 kyr is nearly ubiquitous in these haloes and in most cases leads to the formation of binary or multiple supermassive stellar systems. They also confirm that rapid baryon collapse proceeds for the times required for these stars to form DCBHs. Our simulations suggest that binary or even multiple DCBH formation was the rule rather than the exception in the primordial Universe.

\end{abstract}


\section{Introduction}

Hot, atomically-cooled primordial haloes at $z \sim$ 15 - 20 may have been the birthplaces of the earliest quasars in the universe, more than 200 of which have now been discovered at $z > 6$ \citep[e.g.,][]{fan03}, including nine at $z > 7$ \citep{mort11,ban18,mats19,yang20,wang21}. In this picture, primordial haloes grow to masses of 10$^7$ - 10$^8$ \Ms\ and reach virial temperatures of $\sim$ 10$^4$ K without ever having formed a primordial (Pop III) star, either by being immersed in strong Lyman-Werner (LW) UV backgrounds that destroy all their H$_2$ \citep[e.g.,][]{agarw12,yue14,dfm14,jet14,anna15,anna17} or in highly supersonic baryon streaming motions that delay the collapse of the halo even if H$_2$ is present \citep{th10,greif11,stacy11a,srg17,hir17}. Virial temperatures of 10$^4$ K trigger atomic cooling that leads to catastrophic baryon collapse at infall rates of $\sim$ 0.1 - 1 \Ms\ yr$^{-1}$.  Stellar evolution models predict that such flows, if they persist, would build up cool, red supermassive stars (SMSs) before dying as 100,000 - 300,000 \Ms\ direct-collapse black holes \citep[DCBHs; e.g.,][]{hos13,um16,tyr17,hle18b,tyr21a,herr23a}.  However, it has now been shown that the rare, turbulent haloes that have been shown to form quasars by $z \gtrsim$ 6 in large-scale cosmological simulations \citep{dm12,dm17,lup19,vgf21,lup21} produced DCBHs without UV backgrounds, supersonic baryon streaming motions, or even atomic cooling \citep{latif22b}.

DCBHs are currently the leading candidates for the seeds of the first quasars because ordinary Pop III star BHs are only a few tens to hundreds of solar masses at birth \citep[e.g.,][]{hir13,hir15} and form in low densities that preclude rapid initial growth \citep{wan04,ket04,awa09,wf12,srd18}. In contrast, DCBHs can grow much more rapidly because they are born in dense environments in massive host haloes capable of retaining their fuel supply, even when it is heated by X-rays \citep{jet13}. Runaway stellar collisions in dense, marginally-enriched clusters can create BHs of up to a few thousand solar masses \citep{dv09,latif16,sak17,rein18,boek18} but even these objects may not be massive enough to become quasars by $z >$ 6 (\citealt{smidt18,latif20b,zhu20} -- see also \citealt{titans,maio19}).

Although numerical simulations of rapid baryon collapse in primordial atomically-cooling haloes  have steadily improved over the past decade, they remain a trade-off between resolution and evolution time. The first simulations either resolved sub-AU scales that could capture the formation of the protostar but not the accretion disk around it on parsec (pc) scales \citep{wta08} or larger scales that could follow the evolution of the disk for a few dynamical times but not the SMS at its center \citep{rh09a,rh09b}. These studies found large central infall rates at the onset of collapse but could not determine how long they lasted. Later work at high resolution and somewhat longer evolution times found that these infall rates continued down to scales approaching those of the star but did not run for nearly enough times to follow its evolution \citep{latif13a,rjh14}.  

The implementation of pressure floors \citep{met01} and sink particles at high resolution confirmed that accretion rates could remain large at the smallest scales and persist for a few tens of thousands of years \citep{latif13d,shlos16,chon16}. \citet{chon18} and \citet{rd18b} used sink particles with radiative feedback from the star and a simple prescription for its evolution to follow collapse for 100 kyr and 250 kyr, respectively. They found that radiation from the star was unable to suppress accretion, as did \citet{ard18} and \citet{luo18}, and reported some small-scale fragmentation in the disk at later times \citep[see also][]{bec15,bec18}. \citet{chon18} found that almost half of the fragments in one of their haloes form in binaries that they suspected could later become SMSs, corroborating idealised simulations by \citet{bl03}. However, neither study evolved the disks for long enough times to determine if any of the fragments became stars or were simply subsumed into the central object later on. \citet{suaz19} examined the collapse of atomically cooled haloes at intermediate resolutions in high LW backgrounds for $\sim$ 600 kyr, longer than previous studies but still well short of the formation of a DCBH.

The large inflow rates lasting for millions of years required to form SMSs and DCBHs have only recently been confirmed to occur in numerical simulations \citep{latif20a,ret20}. These models sacrificed the extreme resolution of earlier studies in order to capture the evolution of the gas at the center of the halo over many dynamical times. \citet{latif20a} found that the disk can fragment into binary and multiple SMS systems, but only in a few haloes with high spin parameters that were special cases. Here, we follow the collapse of an ensemble of atomically-cooling haloes over the full range of spin parameters expected for these objects and for a variety of merger histories. We evolve them for up to 3 Myr, enough time for SMS formation and collapse at their centers. For simplicity, we approximate high LW backgrounds by deactivating H$_2$ chemistry in our runs and defer intermediate LW backgrounds with H$_2$ to later studies.  This should be considered to be a strong but useful upper limit to actual LW backgrounds in the primordial Universe because H$_2$ could self-shield in the cores of massive haloes and survive even intense UV fluxes from nearby Pop III stars \citep[][]{sbh10,latif14,latif15a}.  Even small amounts of  H$_2$ could cool gas in the core and result in much less massive stars.  In Section 2 we describe our numerical simulations and discuss halo evolution, disk dynamics and central accretion rates in Section 3. We conclude in Section 4.

\section{Numerical Method}

We first performed a series of low-resolution dark-matter (DM) only runs to identify haloes with a range of assembly histories and spin parameters. These haloes are then resimulated with full baryonic physics at much higher resolution to follow their collapse and the dynamics of the disks at their centers.

\subsection{Enzo}

We use the Enzo adaptive mesh refinement (AMR) cosmology code to model the haloes in our study \citep{enzo}. Enzo has an $N-$body adaptive particle-mesh scheme for evolving DM \citep{efs85,couch91,bn97} that is self-consistently coupled to hydrodynamics and nonequilibrium primordial gas chemistry \citep{anet97,ga08}.  We employ the piecewise parabolic method for gas dynamics \citep[PPM;][]{wc84,bryan95} and apply the HLLC scheme for enhanced stability with strong shocks and rarefaction waves \citep{toro94}. Our simulations include six species (H, He, e$^-$, H$^+$, He$^+$, He$^{2+}$) with no H$_2$ in order to ensure isothermal cooling and collapse.  Our chemistry model includes collisional excitational and ionizational cooling by H and He, recombination cooling, bremsstrahlung cooling, and inverse Compton cooling by the cosmic microwave background (CMB).

\subsection{Simulation Setup}

Haloes of interest are identified from a series of unigrid DM-only simulations in 1.5 $h^{-1}$ Mpc cosmological boxes with a resolution of 256$^3$ and initial conditions generated at $z =$ 200 with MUSIC \citep{hahn11}. We use the second-year \textit{Planck} cosmological parameters:  $\Omega_{\mathrm M} = 0.308$, $\Omega_\Lambda = 0.691$, $\Omega_{\mathrm b}h^2 = 0.0223$, $\sigma_8 =$ 0.816, $h = $ 0.677 and $n =$ 0.968 \citep{planck2}. The Rockstar halo finder \citep{rockstar} is used to find haloes with masses of a few times 10$^7$ \Ms\ in these boxes. We restart our simulations at $z =$ 200 with a top grid resolution of 256$^3$ and three additional nested grids centered on the halo (each with resolution of 256$^3$) which yield effective DM and baryon mass resolutions of 28 \Ms\ and 34 \Ms, respectively. Up to 15 levels of refinement are allowed for a maximum physical resolution of 0.014 pc. A grid is flagged for refinement when baryon or DM overdensities exceed 8.0 times the mean density ($\delta \rho / \bar{\rho} > 8.0$). We resolve the Jeans length by 32 cells to capture the formation of turbulent structures and prevent artificial fragmentation, and we activate refinement on Jeans length at $z =$ 30 \citep{true97,latif13a}. 

DM particles are smoothened at the tenth refinement level, which corresponds to a comoving resolution of 5.72 pc, to avoid spurious effects caused by the discrete sampling of the DM potential.  Below pc scales, collapse is dominated by baryons, and if there is an insufficient number of DM particles at higher resolutions they can unphysically accelerate and heat gas in their vicinity.  Smoothing mitigates such effects.  We also turn on a pressure floor at  the maximum level of refinement in our simulations that prevents collapse on scales below our maximum resolution by ensuring that the cells on that level are Jeans stable.  The pressure floor in our runs smoothens the gravitational potential over 4$\Delta_x$, where $\Delta_x$ is the length of a cell at the highest level of refinement \citep{met01}, so clumps can be smeared out on those scales.   Pressure floors never reverse collapse in our models, just prevent it at the highest level of refinement, and they never result in local expansion of the gas that drives shocks, especially at temperatures of 8000 K.  These points are corroborated by previous simulations of collapse in atomically-cooled haloes with pressure floors and with sink particles without floors that produced essentially identical results \citep{latif13d,latif22b}.  We output the data every 10 kyr during the run and follow the gravitational collapse for about 3 Myr.

\subsection{Halo Selection Criteria}

\begin{figure}
\centering
\includegraphics[width=0.47\textwidth]{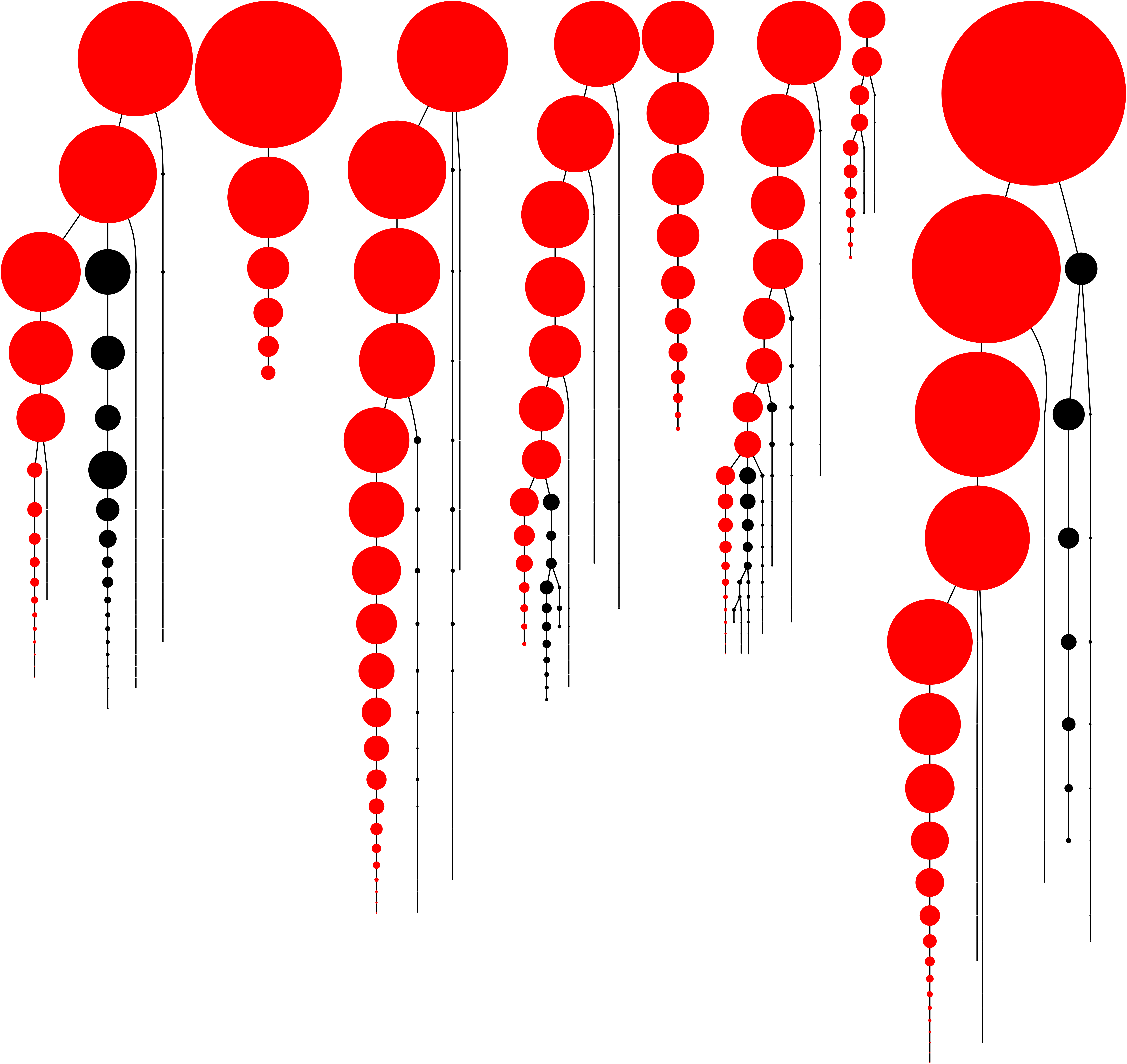}
\caption{Merger histories. Left to right: haloes 1, 2, 8, 10, 12, 16, 19 and 20.}
\label{fig:trees}
\end{figure}

We initially selected 20 haloes from our DM-only runs that reached masses of at least 10$^7$ \Ms\ by $z \sim$ 14 - 20.  From this set we chose only eight haloes for resimulation because of their merger histories and spin parameters at the onset of atomic cooling. They primarily grew through accretion, major mergers or some combination of the two, where a major merger is defined to be the collision of two haloes with a mass ratio of 1/5 or more. They were also chosen so that they spanned the likely range of spin parameters, $\lambda$, for atomically-cooled haloes, where $\lambda$ is \citep{pb69,bul01}
\begin{equation}
\lambda  = \frac{\lvert L \rvert \sqrt{\lvert E \rvert}}{G M^{\frac{5}{2}}} = \frac{\lvert L \rvert}{\sqrt{2 G R M^3}}.
\end{equation}
Here, $L$, $R$, and $M$ are the angular momentum, virial radius and virial mass of the halo, respectively. Values for $\lambda$ for all eight haloes are listed in Table~\ref{tab:haloestats}. The remainder of the 20 haloes were discarded for having merger histories similar to those in our final set.  We save outputs at redshift intervals $\Delta z =$ 0.5 from $z =$ 25 - 15, use Rockstar to locate positions, masses and velocities of the haloes in each redshift bin, and then port them to ytree \citep{ytree} to build merger trees. Assembly histories for these haloes, designated 1, 2, 8, 10, 12, 16, 19 and 20, are shown in Figure~\ref{fig:trees}. The trees extend from $z =$ 25 at the bottom to $z =$ 15 at the top in increments $\Delta z =$ 0.5. Each circle represents an ancestor of the final halo and has a radius proportional to its logarithmically-scaled mass. This proportionality varies between haloes in Figure~\ref{fig:trees} so the same circle radius in different trees does not imply the same mass. The main progenitor line is marked in red.

\begin{table}
\begin{center}
\begin{tabular}{ccccc}
\toprule
halo & $z_{col}$ & mass (\Ms) & spin & mergers \\
\midrule
1    &  16.7  &  3.68e7  &  0.0389  &  1  \\
2    &  14.5  &  8.47e7  &  0.0388  &  0  \\
8    &  20.4  &  1.15e7  &  0.0321  &  0  \\
10  &  17.3  &  2.60e7  &  0.0500  &  1  \\
12  &  16.8  &  1.93e7  &  0.0471  &  0  \\
16  &  16.5  &  2.91e7  &  0.0258  &  3  \\
19  &  13.9  &  2.12e7  &  0.0072  &  0  \\
20  &  17.7  &  3.56e7  &  0.0199  &  0  \\
\bottomrule
\end{tabular}
\end{center}
\caption{Halo properties at the onset of atomic cooling. From left to right: halo, collapse redshift, mass at the onset of atomic cooling, spin parameter and number of major mergers.}
\label{tab:haloestats}
\end{table}

Halo 12 is an example of a halo that has grown primarily by accretion while halo 16 has three major mergers. The others lie somewhere between these two extremes. Most of the haloes have lines of low-mass haloes in their ancestry, visible as the long vertical black tails with small black dots along them. Halo 1 grows slowly at early times but then suddenly rises in mass at $z =$ 17.5 because of a surge in accretion and a merger. Halo 2 has the highest accretion rate of the sample but does not begin to cool before reaching a mass of 8.47 $\times$ 10$^7$ \Ms\ at $z =$ 14.5, likely because of turbulence driven by rapid infall. Halo 8 begins to cool at the earliest times at a mass of 1.15 $\times$ 10$^7$ \Ms\ at $z =$ 20.4. 

\begin{figure*}
\begin{center}
\begin{tabular}{cc}
\includegraphics[width=0.45\textwidth]{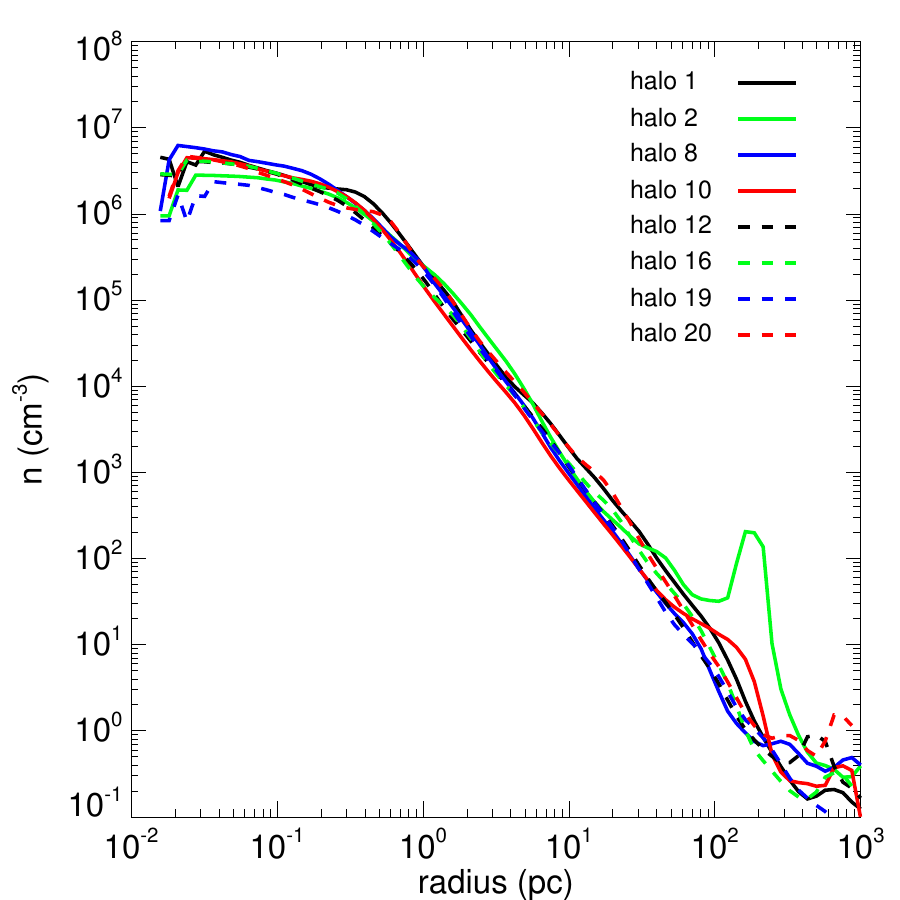} &
\includegraphics[width=0.45\textwidth]{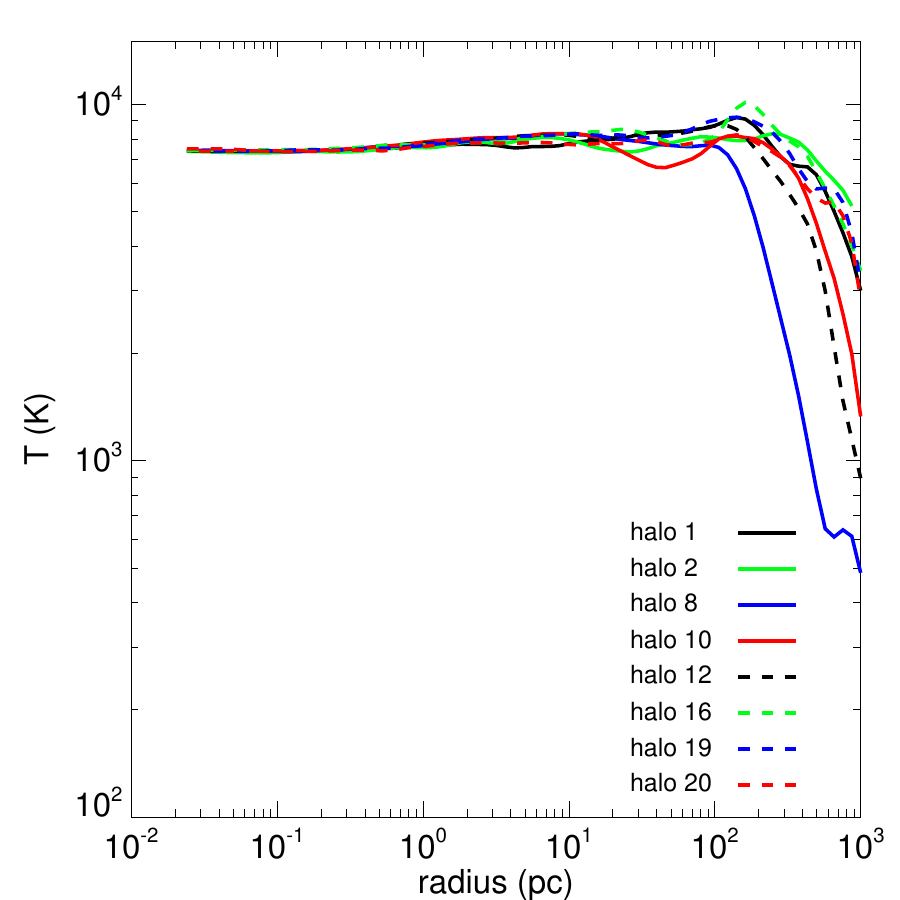} \\
\end{tabular}
\end{center}
\caption{Spherically-averaged density (left) and temperature (right) profiles of the eight haloes at the onset of atomic cooling and rapid baryon collapse.}
\label{fig:1Dprof}
\end{figure*}

Halo 10 has a major merger early in its history and has the highest spin parameter at the time it begins to cool at $z =$ 17.3 at a mass of 2.6 $\times$ 10$^7$ \Ms. Halo 12 has the next highest spin when it begins to cool at $z =$ 16.8 at a mass of 1.93 $\times$ 10$^7$ \Ms, which is interesting because it is almost entirely the product of accretion, not mergers that could have spun it up by tidal torquing. Halo 16 undergoes three major mergers at $z =$ 18.5, 17.5, and 16.5 and minor mergers at z=18.5 and 15. It has the lowest accretion rates in our sample.  Halo 19 begins to cool at the lowest redshift, $z =$ 13.9 at a mass of 2.12 $\times$ 10$^7$ \Ms, and has the lowest spin at the onset of cooling.  Like halo 12, which has the second highest spin parameter, it too is mostly the result of accretion, not mergers.  Halo 20, which has the second lowest spin parameter, is also mostly built up by accretion but has a number of minor mergers in its history. We summarise these results in Table~\ref{tab:haloestats}. In sum, haloes 1, 10 and 16 have undergone major mergers during their formation, haloes 2, 12 and 19 have grown mainly via accretion, and haloes 8 and 20 are products of both minor mergers and accretion. Redshifts at the onset of atomic cooling are determined in our runs with both gas and DM for accuracy.

\section{Results}

Rapid baryon collapse triggered by atomic cooling leads to the formation of large, rotationally-supported disks at the centers of all eight haloes.  We now discuss the evolution of these disks, their accretion rates, and their stability over time.

\subsection{Initial Baryon Collapse}

We show spherically-averaged density and temperature profiles for our haloes at the onset of cooling and collapse in Figure~\ref{fig:1Dprof}. The densities all closely approximate the Larson-Penston solution, $\rho \propto r^{-2.2}$, which slightly deviates from the classic $r^{-2}$ density profile expected for self-gravitating isothermal spheres because of the presence of DM in the halo. The virial shock of the halo is visible as the bump in density and temperature at 200 pc - 800 pc in the profiles. Densities range from 10$^6$ - 10$^7$ cm$^{-3}$ in the core to 0.1 - 1 cm$^{-3}$ at the virial shock. They level off at small radii because of the absence of torques within the Jeans sphere (not because of pressure floors, which are imposed just after this stage of collapse). The large bump in density at $\sim$ 200 pc in halo 2 is due to another halo that is about to merge with it.

The temperature profiles show that collapse due to atomic cooling is nearly isothermal, with flat $\sim$ 8000 K temperatures extending out to $\sim$ 100 pc. Gas falling onto the halo is first heated to 10,000 K as it passes through the virial shock and then cools to $\sim$ 8000 K as it settles deeper into the gravitational potential well of the halo. Because of our resolution limits, gas never reaches densities where it becomes optically thick to Ly$\alpha$ photons \citep[e.g.,][]{aaron17,sb19}. At higher levels of refinement these photons could become partly trapped by frequent resonant scatterings, reducing cooling efficiencies and increasing temperatures deep in the core. We also note that at number densities $n \gtrsim$ 10$^8$ cm$^{-3}$ H$^-$ cooling becomes important and would reduce gas temperatures to $\sim$ 5000 K before the gas becomes optically thick to Ly$\alpha$, further increasing densities and potentially trapping more cooling photons at later times.

\subsection{Disk Formation / Evolution}
\label{sec:disks}

\begin{figure*}
\begin{center}
\begin{tabular}{cccc}
\includegraphics[width=0.22\textwidth]{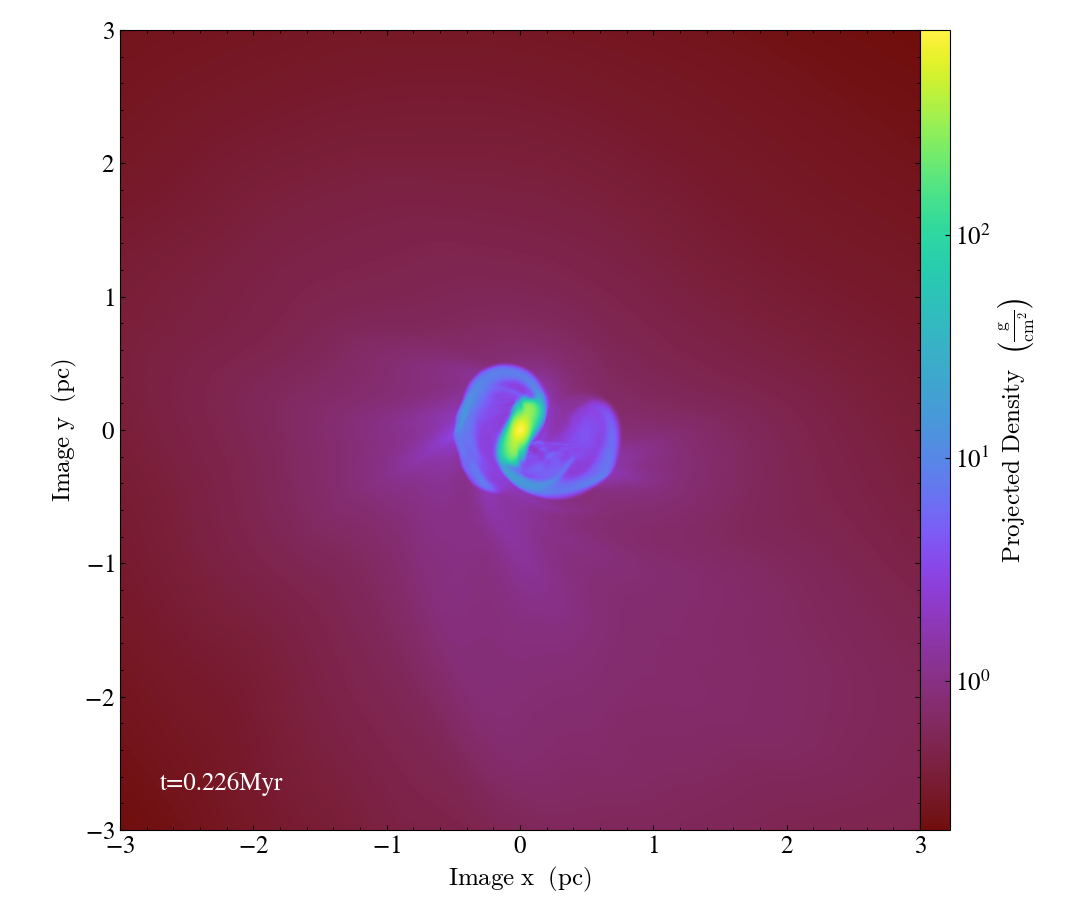} &
\includegraphics[width=0.22\textwidth]{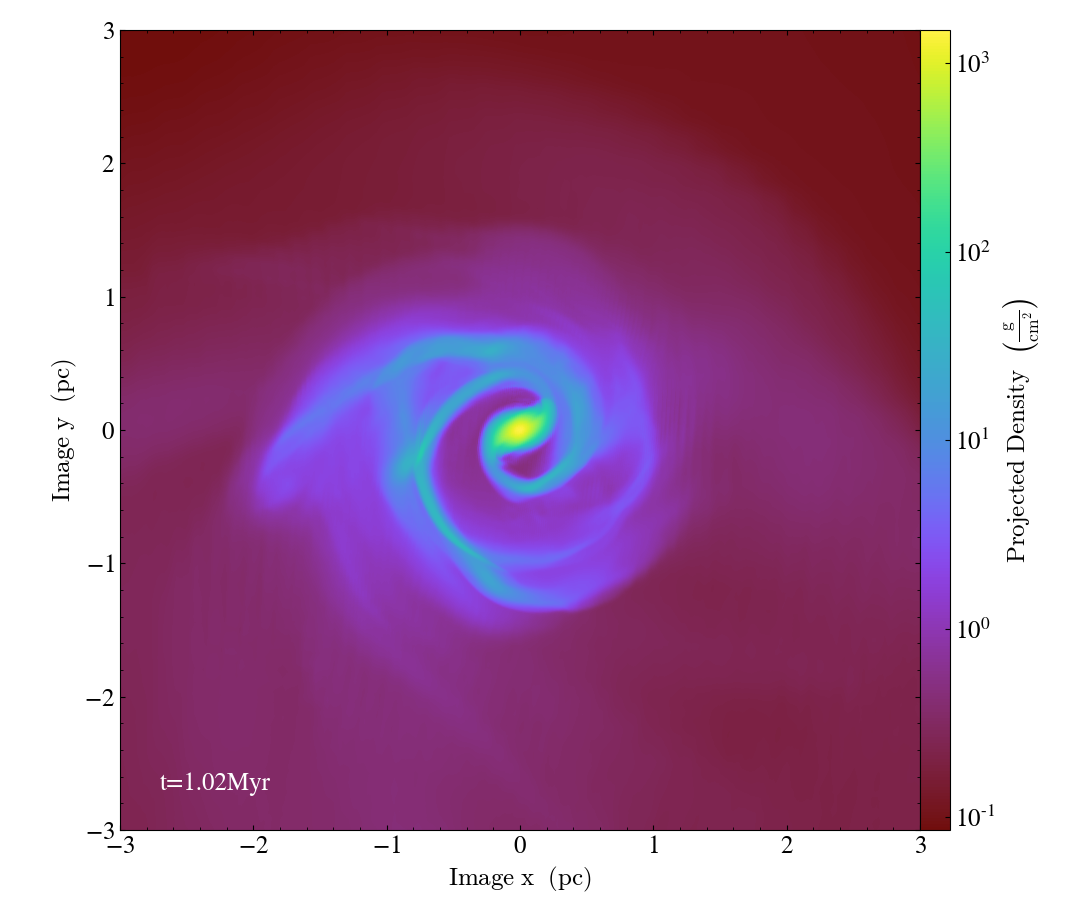} &
\includegraphics[width=0.22\textwidth]{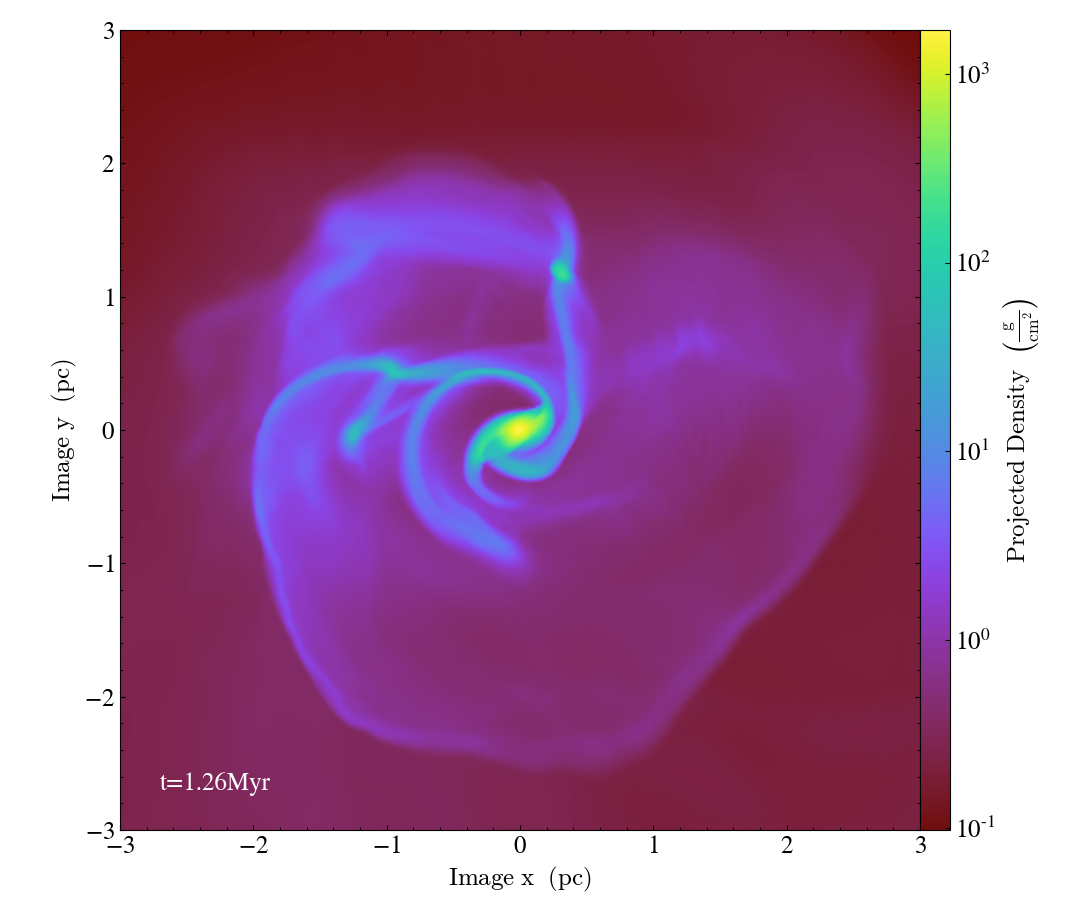} &
\includegraphics[width=0.22\textwidth]{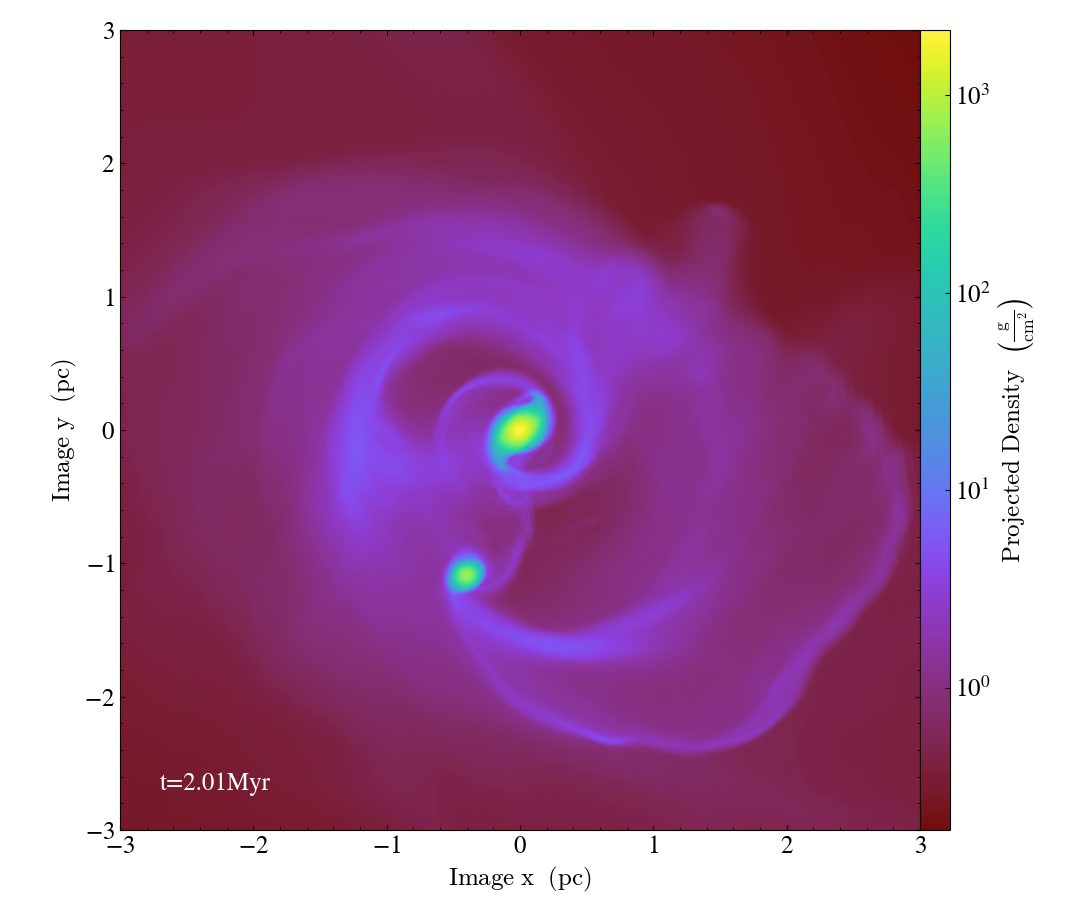} \\
\includegraphics[width=0.22\textwidth]{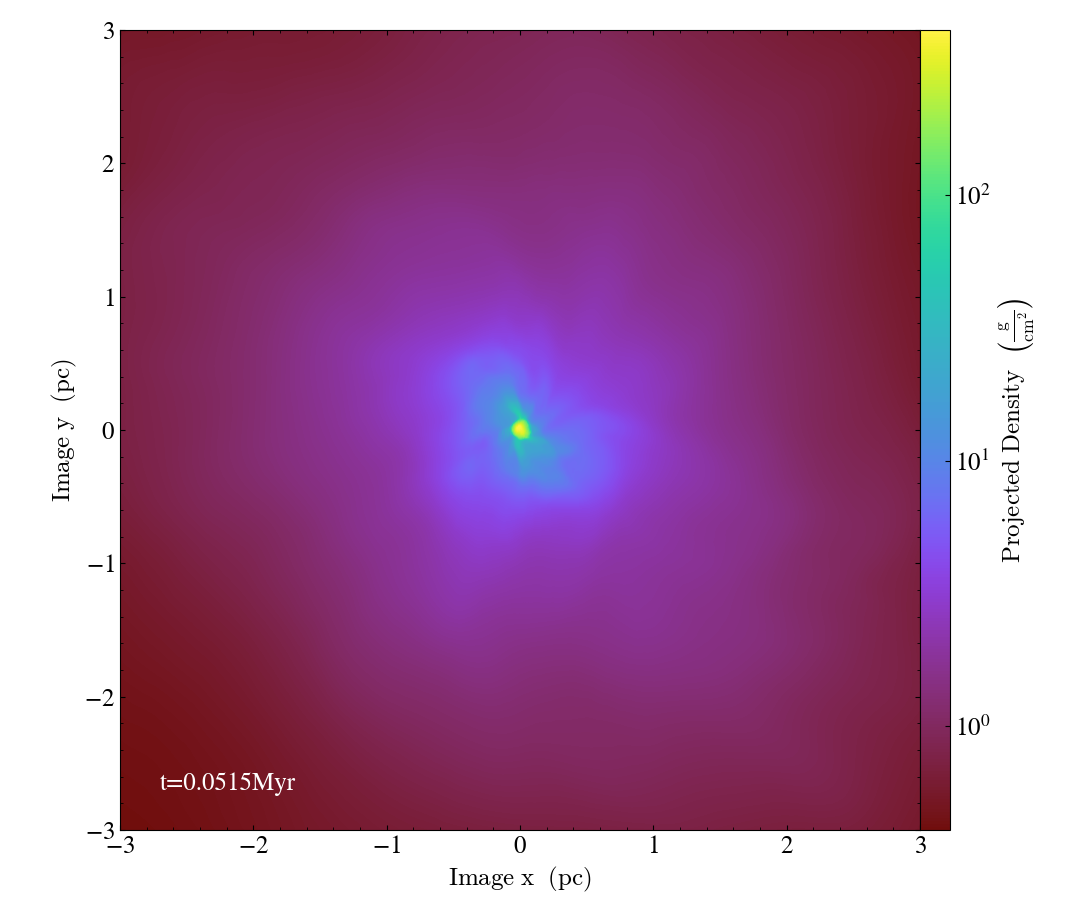} &
\includegraphics[width=0.22\textwidth]{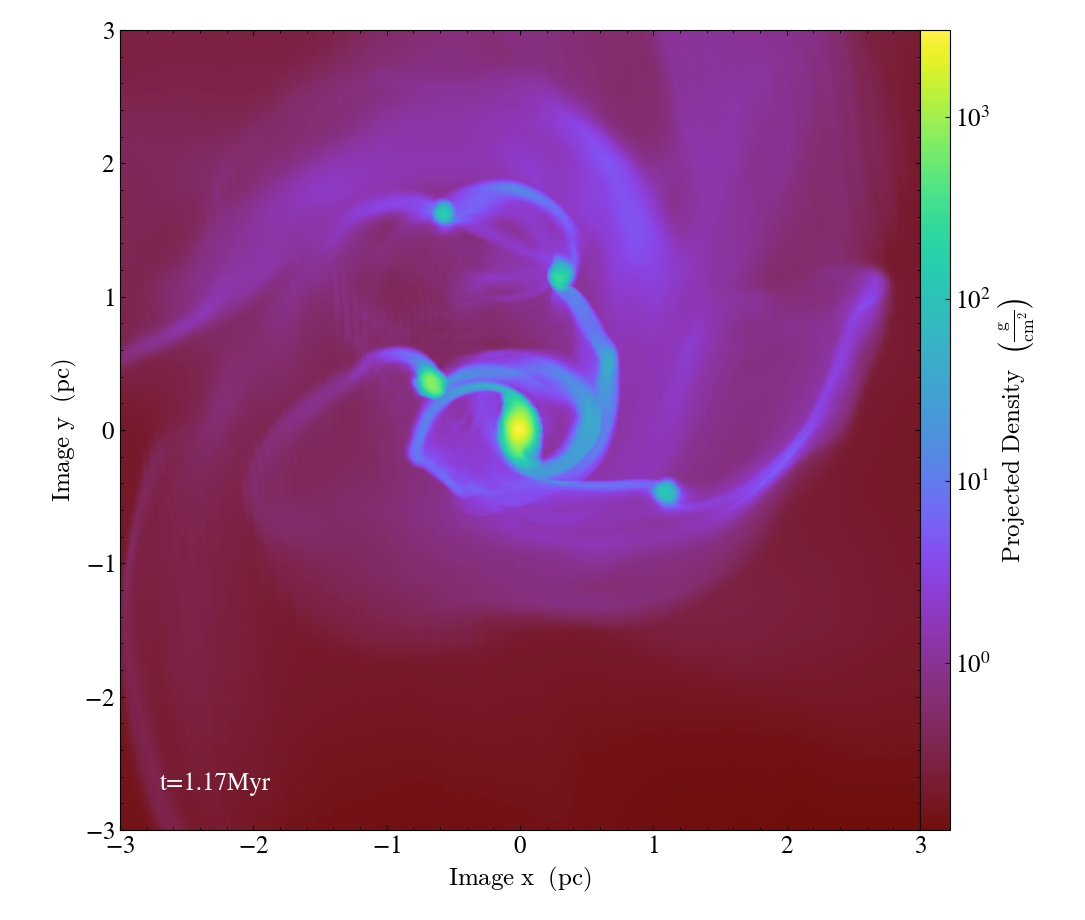} &
\includegraphics[width=0.22\textwidth]{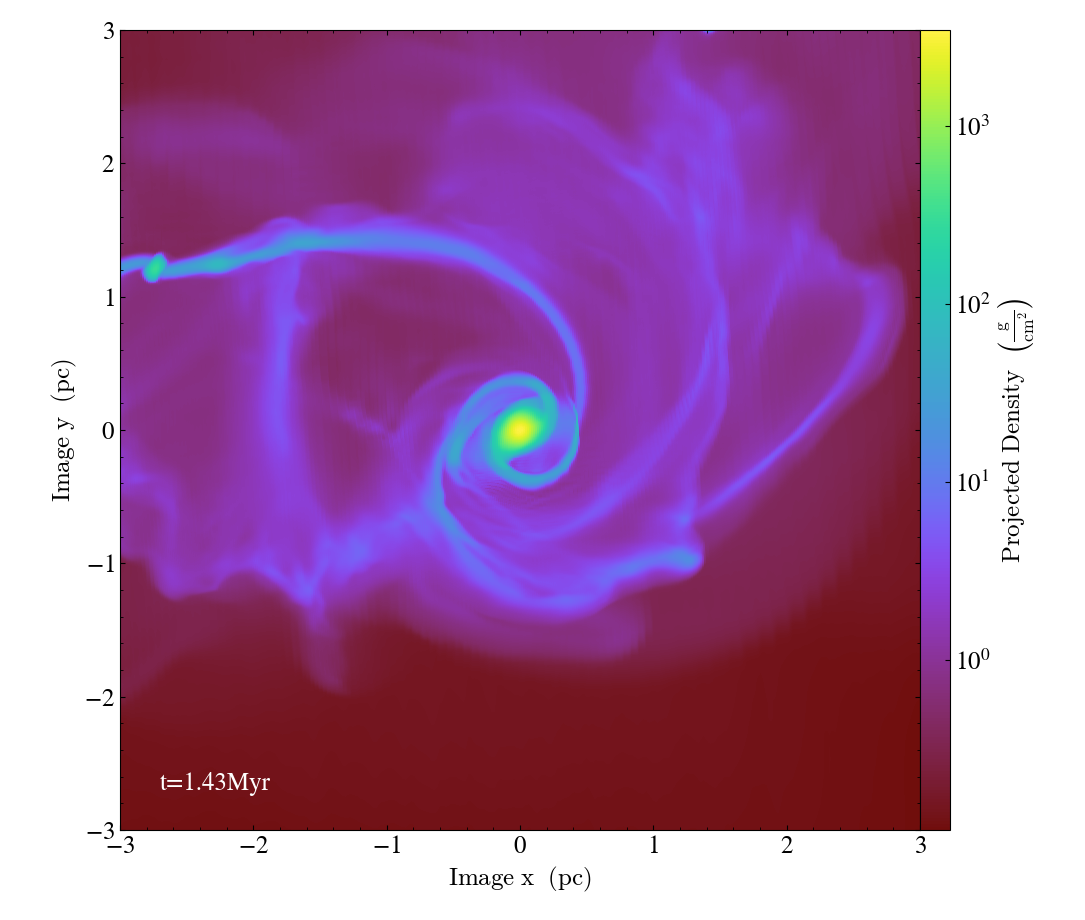} &
\includegraphics[width=0.22\textwidth]{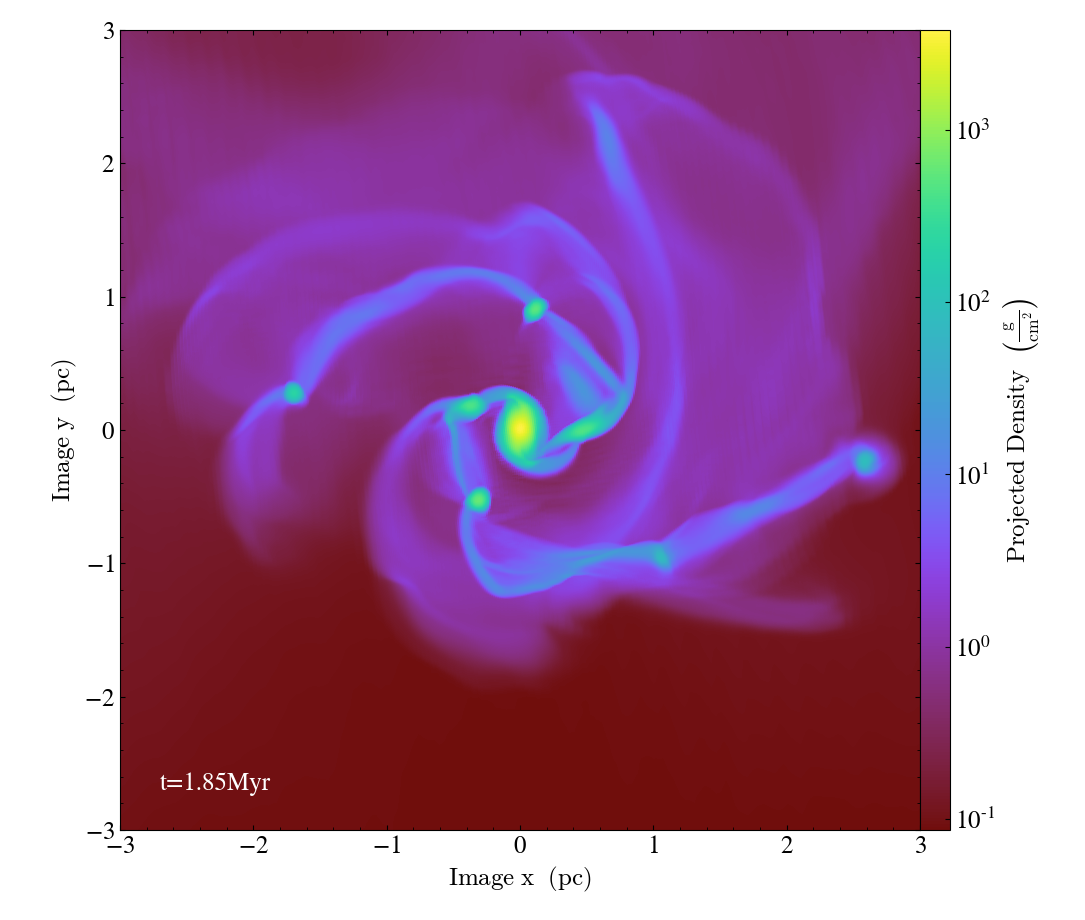} \\
\includegraphics[width=0.22\textwidth]{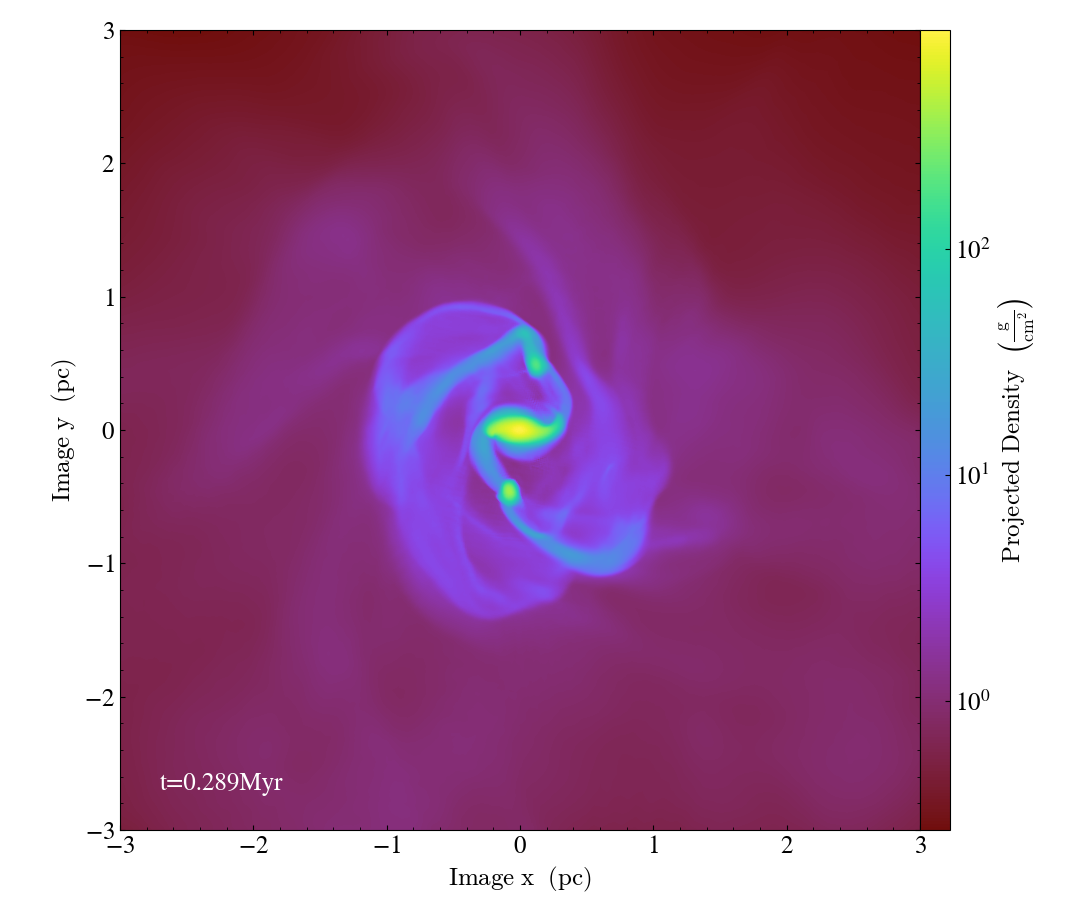} &
\includegraphics[width=0.22\textwidth]{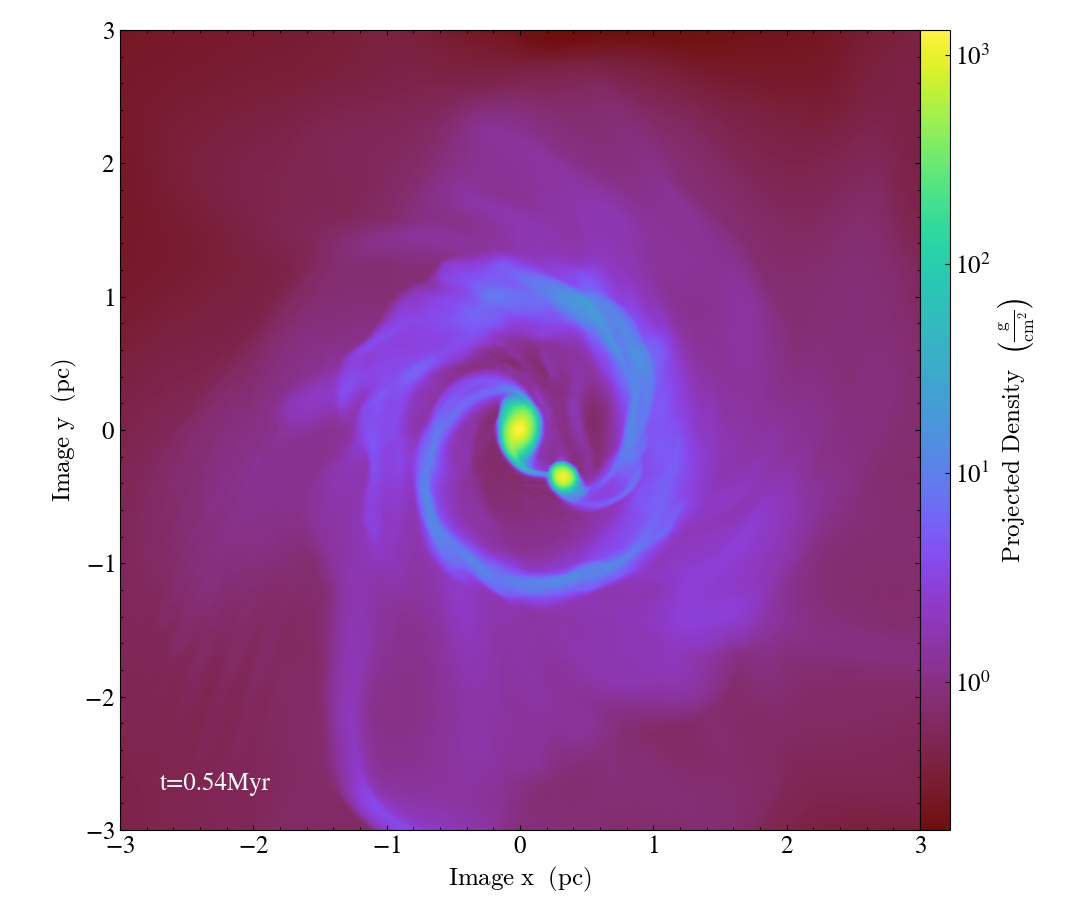} &
\includegraphics[width=0.22\textwidth]{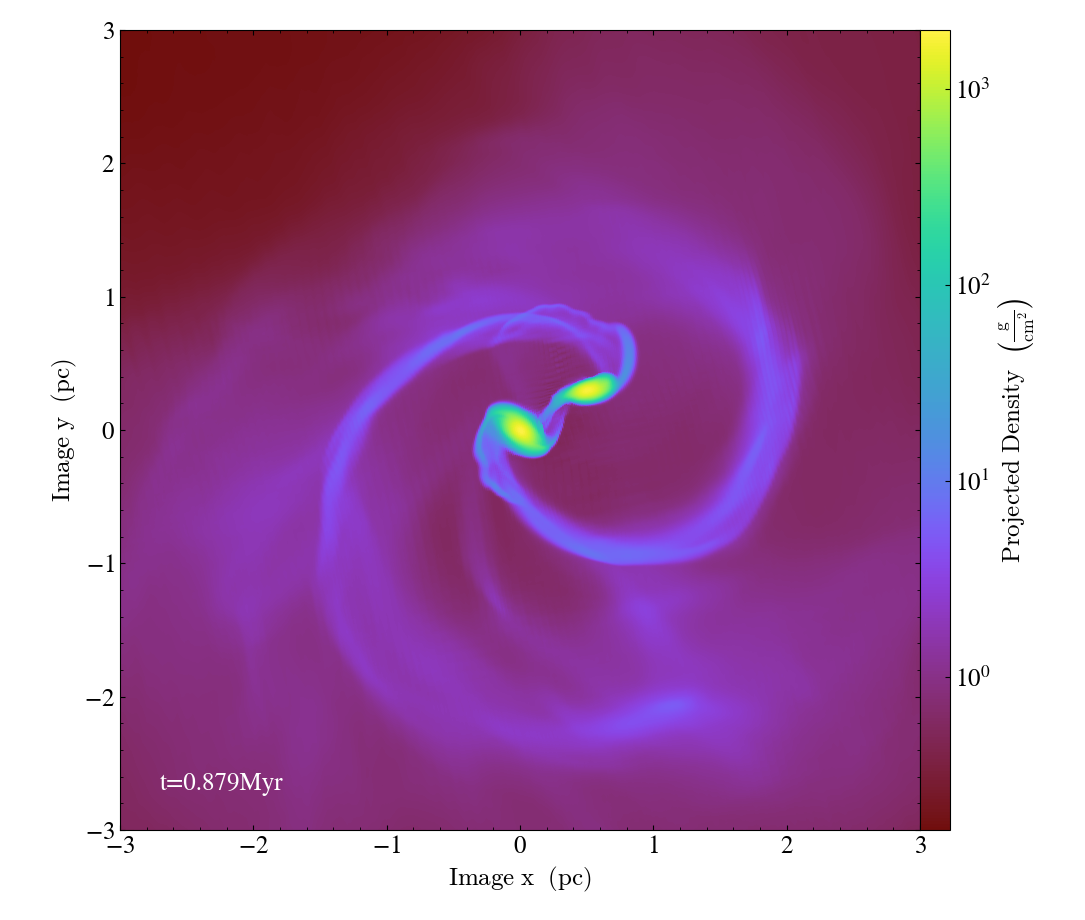} &
\includegraphics[width=0.22\textwidth]{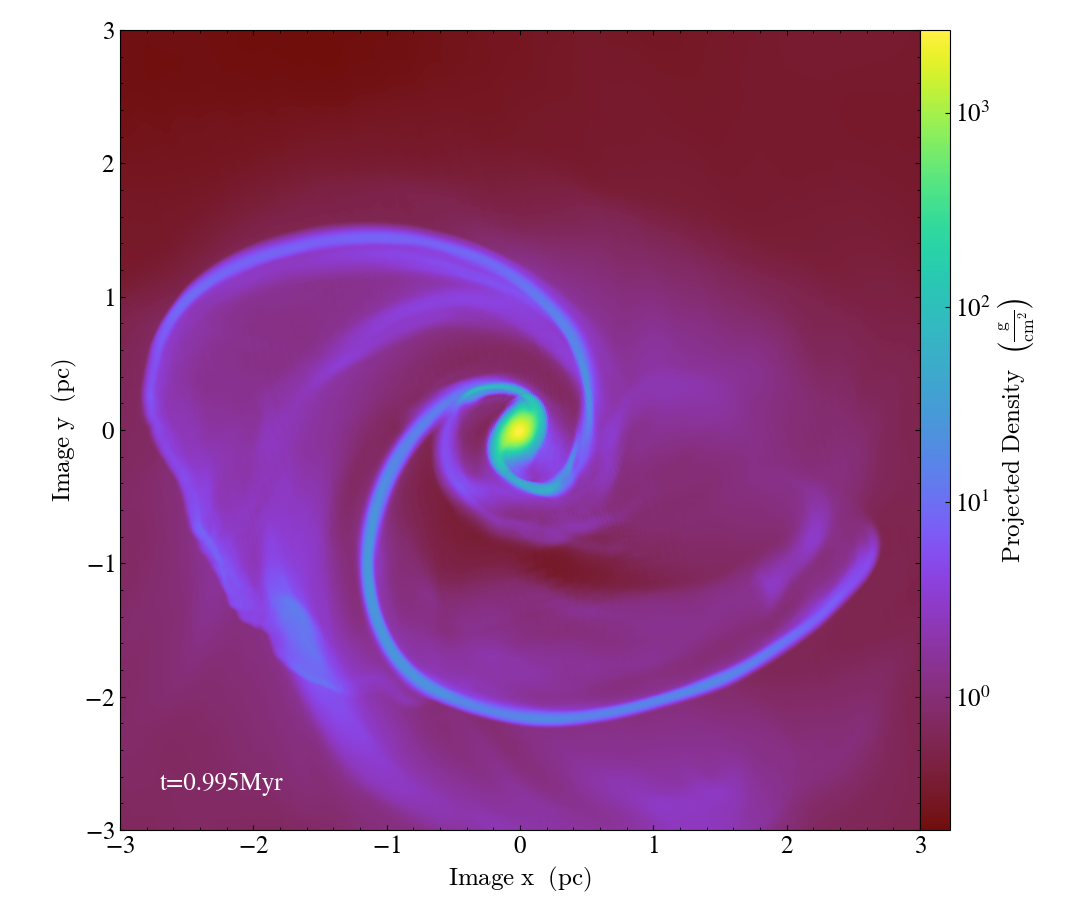} \\
\end{tabular}
\end{center}
\caption{Projections of gas density in 3 of the 8 disks in our simulations.  Top row: the relatively stable disk in halo 10. Left to right:  226 kyr, 1.02 Myr, 1.26 Myr, and 2.01 Myr. Center row: the more turbulent, rapidly fragmenting disk in halo 12. Left to right:  51 kyr, 1.17 Myr, 1.43 Myr and 1.85 Myr.  Bottom row: early binary disk formation in halo 1. Left to right:  289 kyr, 540 kyr, 879 kyr and 995 kyr.}
\label{fig:evol}
\end{figure*}

Rapid baryon collapse leads first to the formation of a dense core and then a rotationally-supported, self-gravitating disk with an initial radius of $\sim$ 0.2 pc by $\sim$ 200 kyr. A bar instability then appears and creates two spiral arms.  The disk thereafter can exhibit a range of morphologies, as we discuss below, but continues to grow in mass and radius and generally becomes more turbulent over time. This turbulence increases the effective alpha parameter of the disk and it grows to over 1 pc in radius in all eight haloes. 

Density projections of the disks in haloes 10, 12 and 1 are shown in Figure~\ref{fig:evol}. They represent the range of disk evolution in our runs:  relatively quiescent ones, as in halo 10; more turbulent disks that undergo multiple episodes of fragmentation, as in halo 12; and the prompt formation of interacting binary disks at early times, as in halo 1. In the initial stages of disk formation the collapse of low-angular momentum gas creates the small, dense cores like the one at 51 kyr in halo 12. They are soon followed by the appearance of barred spiral arms as gas at higher angular momentum rains down onto the disk, as shown at 226 kyr in halo 10. Since radiation from the star cannot prevent gas from continuing to fall onto the disk at high rates, it grows rapidly in radius, in some cases to $\sim$ 3 pc by the end of the runs. Neither the bars nor viscous processes can transport angular momentum out from the center of the disk as fast as gas rains down on it so its spiral arms cannot maintain a single angular velocity and they begin to fracture into clumps, generally by 500 - 700 kyr but as early as $\sim$ 250 kyr in halo 1.  Disk fragmentation thus proceeds here via the spiral arm instability (SAI) mechanism discussed in \citet{iy20} in which spiral arms form first and then become locally unstable, overcoming the stabilizing effects such as gas pressure and Coriolis force. Most fragments spiral into the center of the disk but some persist for times that are sufficient to form satellite disks, as we discuss below. Tidal torquing between fragments and the disk often cause the center of the disk to eject tidal tails when the clumps crash into it. Turbulence in the disks also breaks up spiral arms to some degree.

The disk in halo 10 is the most stable one and does not begin to fragment until 1.14 Myr, when it has grown to $\sim$ 2 pc in radius and become more turbulent. It thereafter forms a number of clumps, but they are quickly torn apart by gravitational torques and taken back up into the spiral arms except for one that forms at 1.26 Myr about 1.2 pc from the center. This clump grows in mass over time, collapsing into a second, smaller disk that is in a highly elliptical orbit with the first.  This second disk exchanges mass with the first in a series of close encounters by the end of the run at 2.01 Myr. Because the second disk appears well after the first one, either a DCBH - SMS binary system or a binary DCBH could form after the second star collapses.

\begin{figure*}
\begin{center}
\begin{tabular}{cc}
\includegraphics[width=0.45\textwidth]{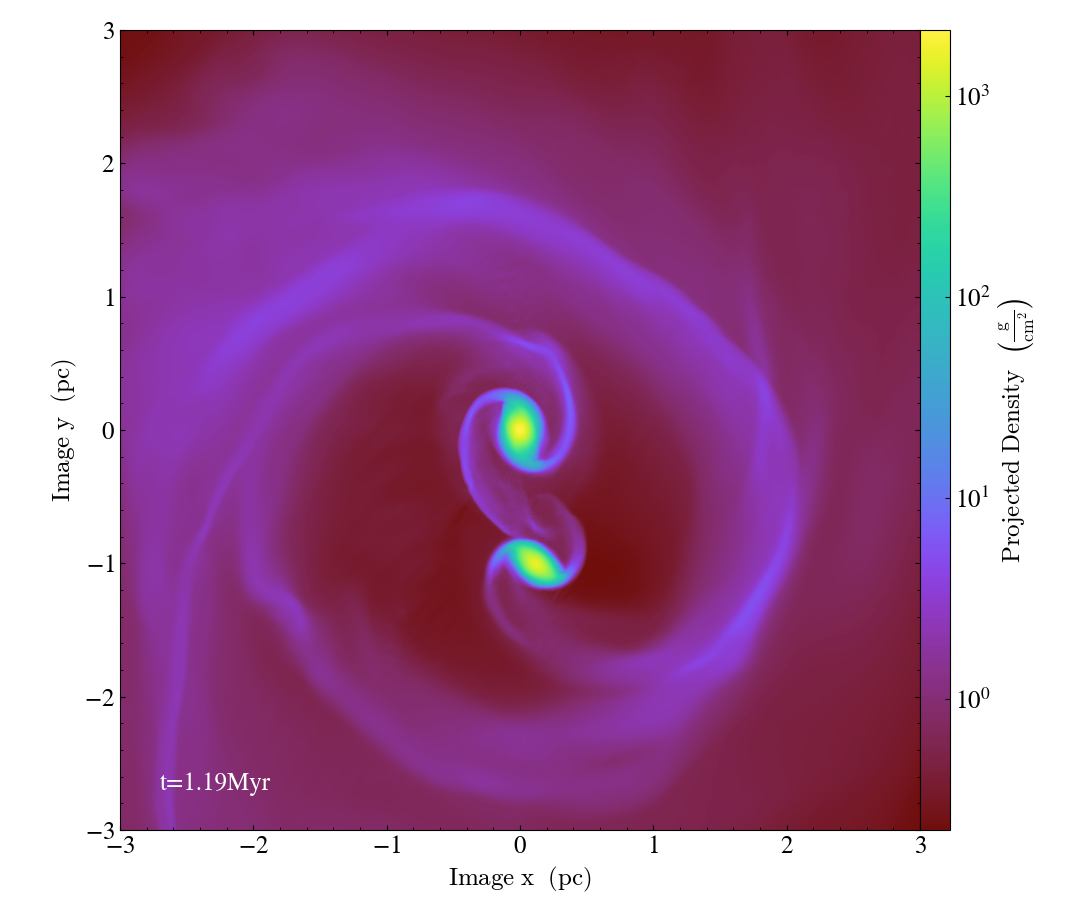} &
\includegraphics[width=0.45\textwidth]{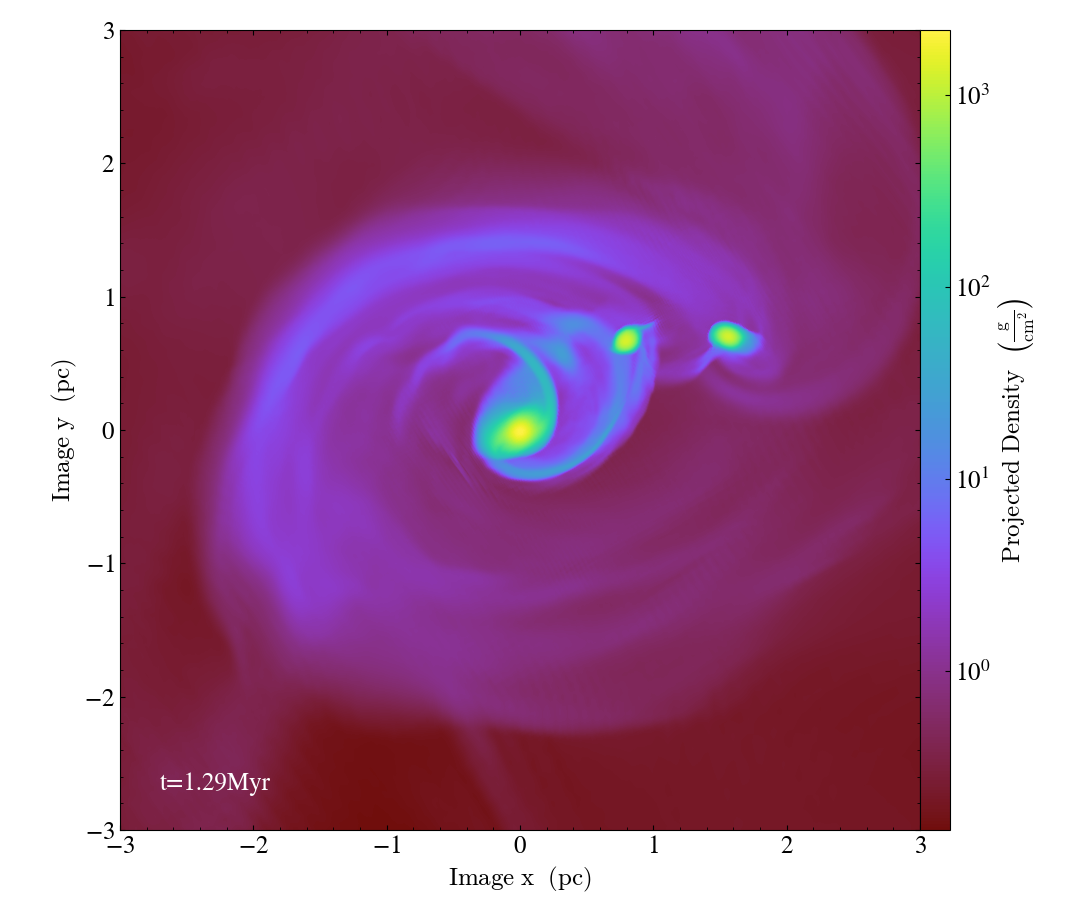} \\
\end{tabular}
\end{center}
\caption{Binary and multiple disk systems in our runs.  Left:  the main disk and its satellite in halo 20 at 1.19 Myr.  Right:  Two smaller disks orbiting the main disk at 1.29 Myr in halo 2.}
\label{fig:mdisks}
\end{figure*}

There is more turbulence in the disk in halo 12 at early times and it produces two episodes of violent fragmentation, shown at 1.17 Myr and 1.85 Myr in Figure~\ref{fig:evol}. Between episodes, most of the short-lived clumps are torn apart by gravitational torques and ejected in the tidal streams visible at 1.43 Myr, but one clump flattens into a secondary disk that persists from 889 kyr - 1.39 Myr. Consequently, it is likely that a second SMS will form in this system, as discussed below. The star at the center of the main disk likely collapses to a DCBH before the second episode of clumping, so some of the clumps would rapidly fuel the growth of the BH when they crash into the center of the disk.  

The disk in halo 1 exhibits morphologies similar to the others at first but begins to fragment at early times, 289 kyr after formation. This episode is short-lived, however, because of the formation of a second, satellite disk at 366 kyr at an initial distance of $\sim$ 1 pc from its more massive companion. It orbits its companion until being torn apart and subsumed into the main disk at 947 kyr, which results in the ejection of the multiple tidal tails shown in Figure~\ref{fig:evol} at 995 kyr. The orbit is highly eccentric and precesses around the main disk, with disk-disk separations varying from $\sim$ 0.5 - 1.2 pc. Mass exchange between the two disks occurs a number of times during close encounters, as shown at 879 kyr.  Tidal torques between the two disks suppress subsequent fragmentation in both, with only two short-lived clumps forming after the appearance of the second disk (fragmentation can also be pre-empted as the satellite disk sweeps up gas along its orbit). The survival of the second disk for $\sim$ 600 kyr leads to the likely formation of a second SMS that is later drawn into a close binary with the star in the main disk, with final separations of less than 0.1 pc at the time they collapse to DCBHs. Large accretion rates in both disks corroborate the likely eventual formation of a binary DCBH, as discussed below.

\subsection{Binary and Multiple Disk Systems}

Although we noted the appearance of binary accretion disks in haloes 1, 10 and 12 in Section~\ref{sec:disks}, they form in all the haloes and in some cases multiple disks appear. For example, three additional disks appear in halo 2: from 666 kyr - 1.04 Myr, 937 kyr - 1.48 Myr and 1.18 Myr - 1.32 Myr (right panel of Figure~\ref{fig:mdisks}). There is almost no overlap in time between the first satellite disk and the other two, which coevolve and even briefly exchange mass with each other at one point. In halo 8 just one binary disk appears in the run from 685 kyr - 1.0 Myr. In halo 16, which hosts a relatively stable disk like that in halo 10, a fragment forms at 1.75 Myr, is flung into a highly elliptical orbit by three-body interactions with the main disk and another clump at 2.01 Myr, and then flattens into a satellite accretion disk by 2.7 Myr that is still orbiting the original disk at the end of the simulation at 3.05 Myr. Two additional disks form in halo 19, from 0.675 kyr - 907 kyr and 907 kyr - 1.41 Myr. Finally, in halo 20, two disk fragments collide at 639 kyr and form a dense clump that flattens into a disk by 774 kyr.  It grows and soon rivals the original disk in mass (see the left panel of Figure~\ref{fig:mdisks}) but is then almost destroyed in close encounters with this disk twice, at 1.36 Myr and at 1.61 Myr. In both cases tidal streams are ejected from the site of the interaction and the satellite disk retains only a small fraction of its mass after 1.61 Myr.

\begin{figure}
\centering
\includegraphics[width=0.47\textwidth]{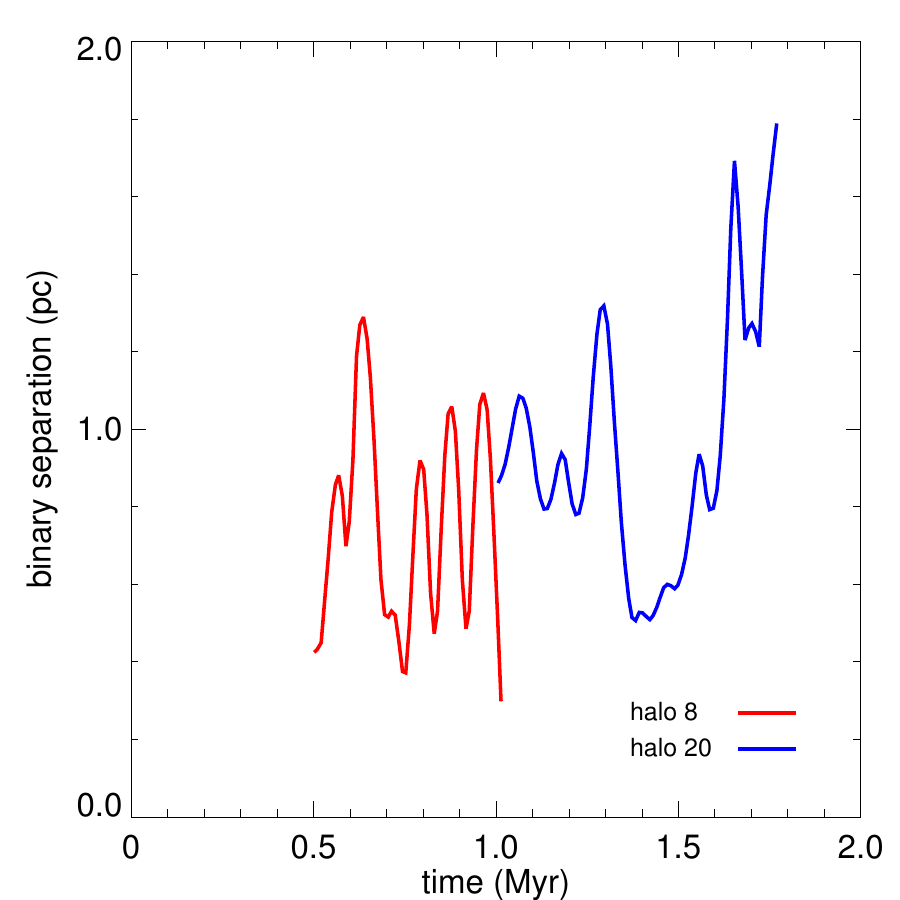}
\caption{Center-to-center separations between the main disk and satellite disk in haloes 8 and 20.}
\label{fig:bsep}
\end{figure}

We show center-to-center separations between the main disk and its satellite over time in haloes 8 and 20 in Figure~\ref{fig:bsep}. Orbits vary from fairly regular, as in halo 8, to highly irregular, as in halo 20. Binary motion in the other haloes falls somewhere in between these two extremes. Separation distances vary from $\sim$ 0.3 pc - 2 pc, and the satellite disks typically make 5 - 10 orbits before either crashing into the center of the main disk or being torn apart by gravitational torques.

The binary disks in our models live for shorter times than those in \citet{latif20a}. Their disks survive for longer times because their simulations probed the upper extremes in spin parameter for the host haloes.  Their large angular momenta allowed the satellite disks to orbit the main disks for longer times before crashing into them. In contrast, the haloes in our study have a range of spin parameters that do not reach their upper or lower extremes so satellite disks are taken up into the main disks in less time. 

\begin{figure*} 
\begin{center}
\begin{tabular}{ccccc}
\epsfig{file=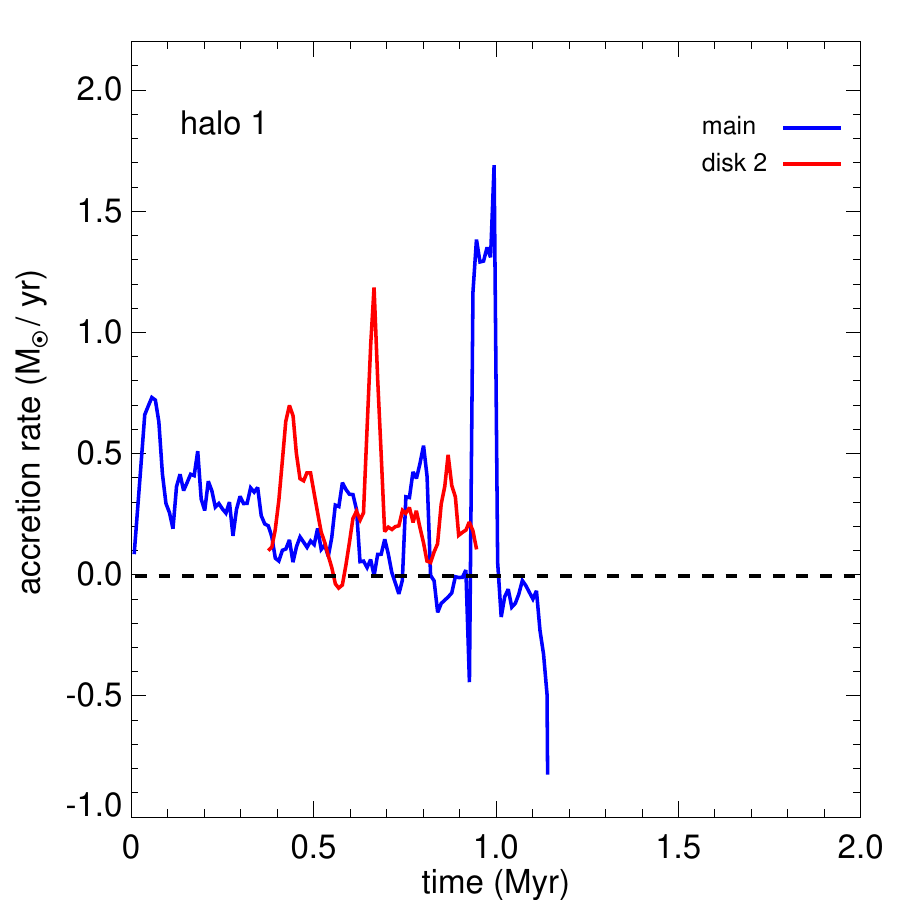,width=0.32\linewidth,clip=}  &  &   &  &
\epsfig{file=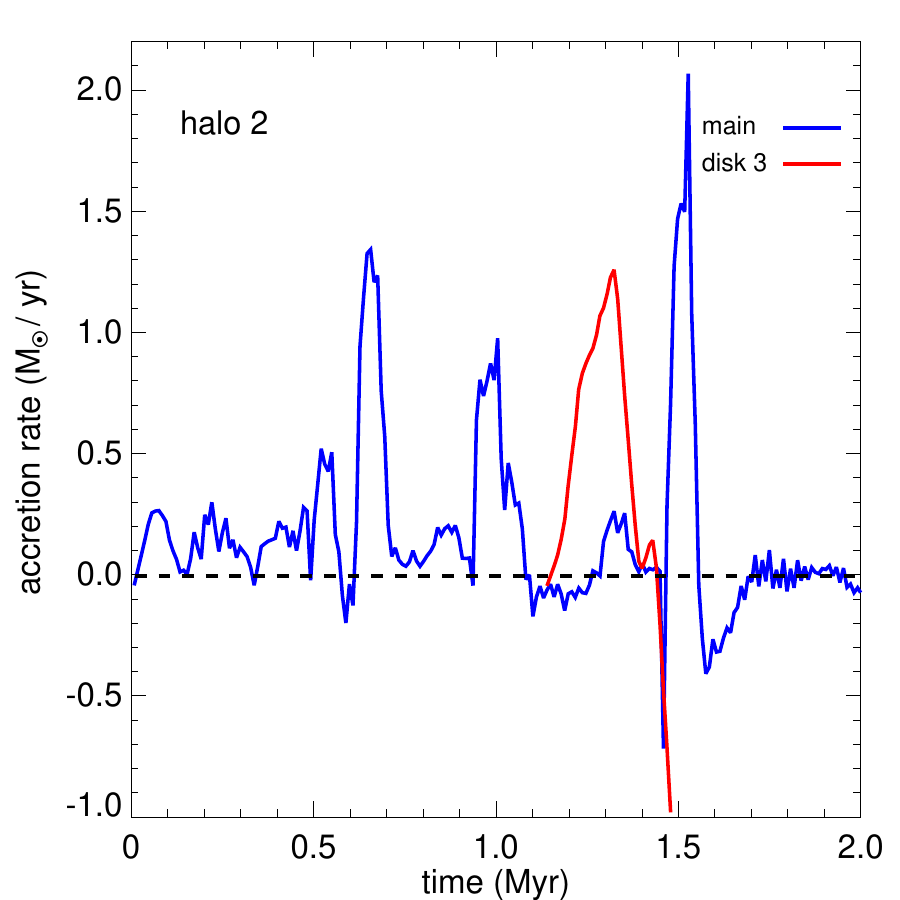,width=0.32\linewidth,clip=}  \\
\epsfig{file=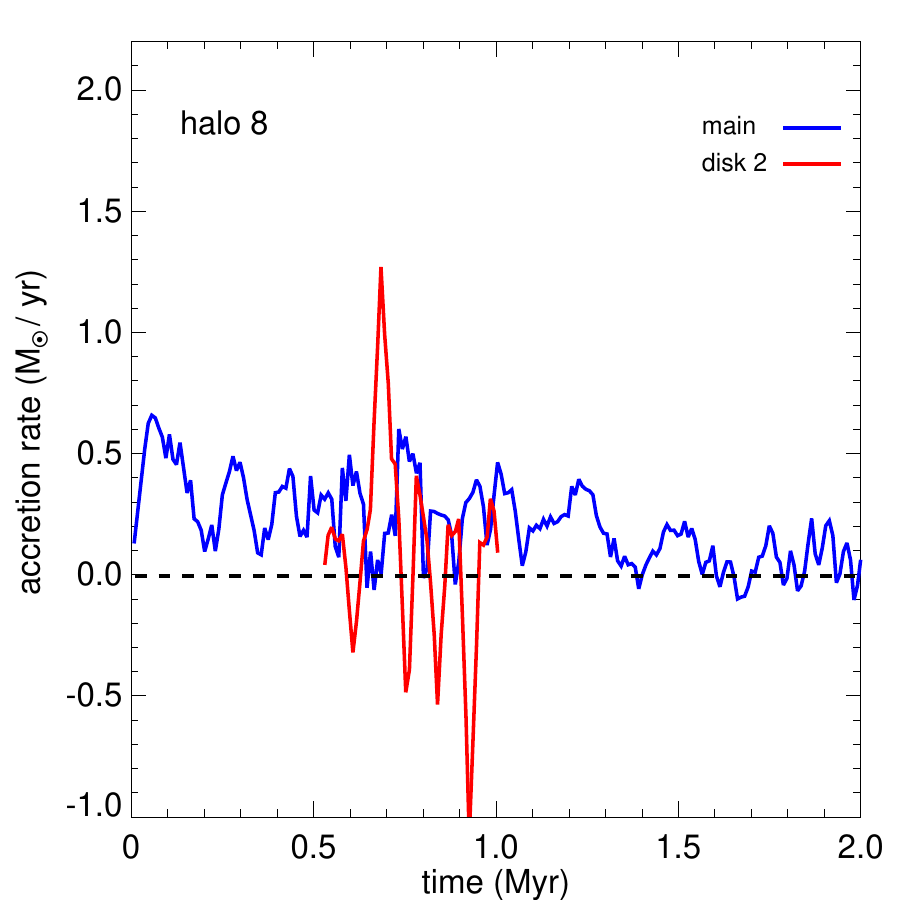,width=0.32\linewidth,clip=}  &  &   &  &
\epsfig{file=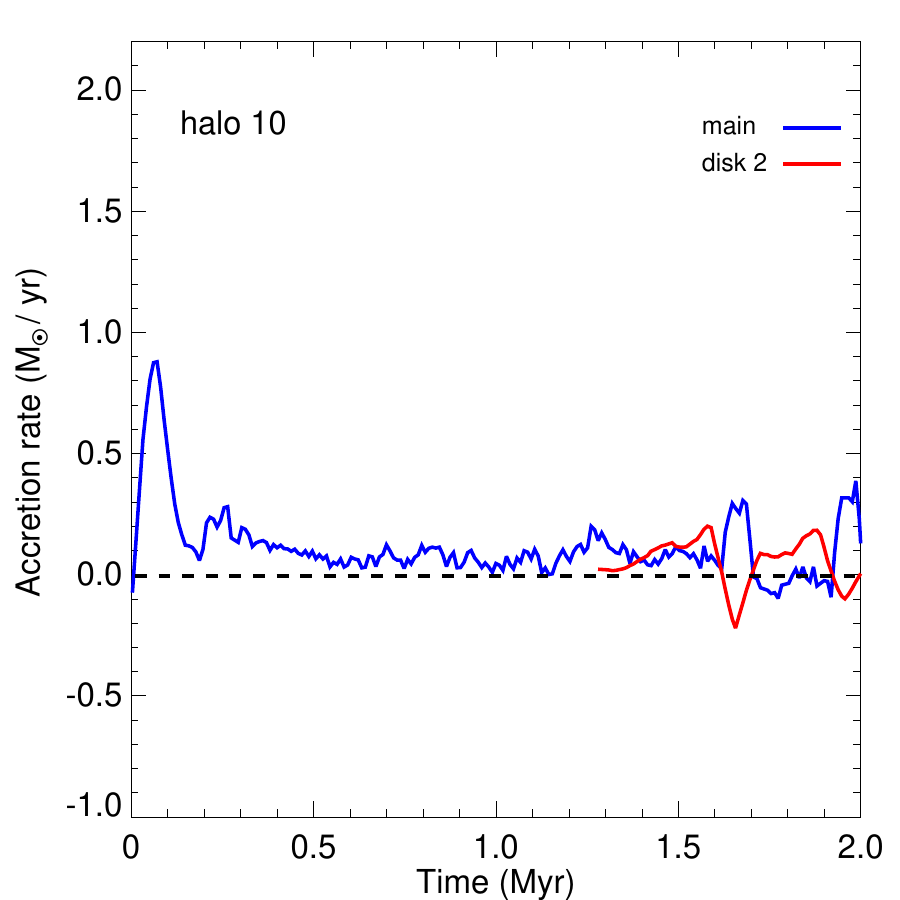,width=0.32\linewidth,clip=}  \\
\epsfig{file=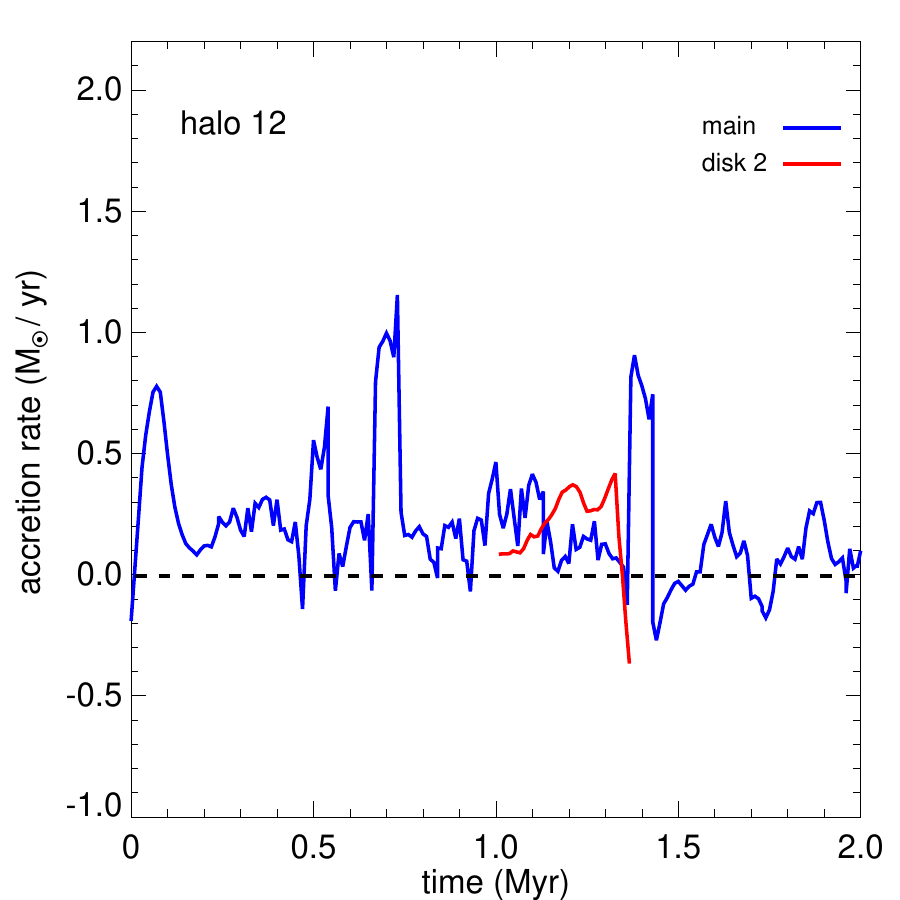,width=0.32\linewidth,clip=}  &  &   &  &
\epsfig{file=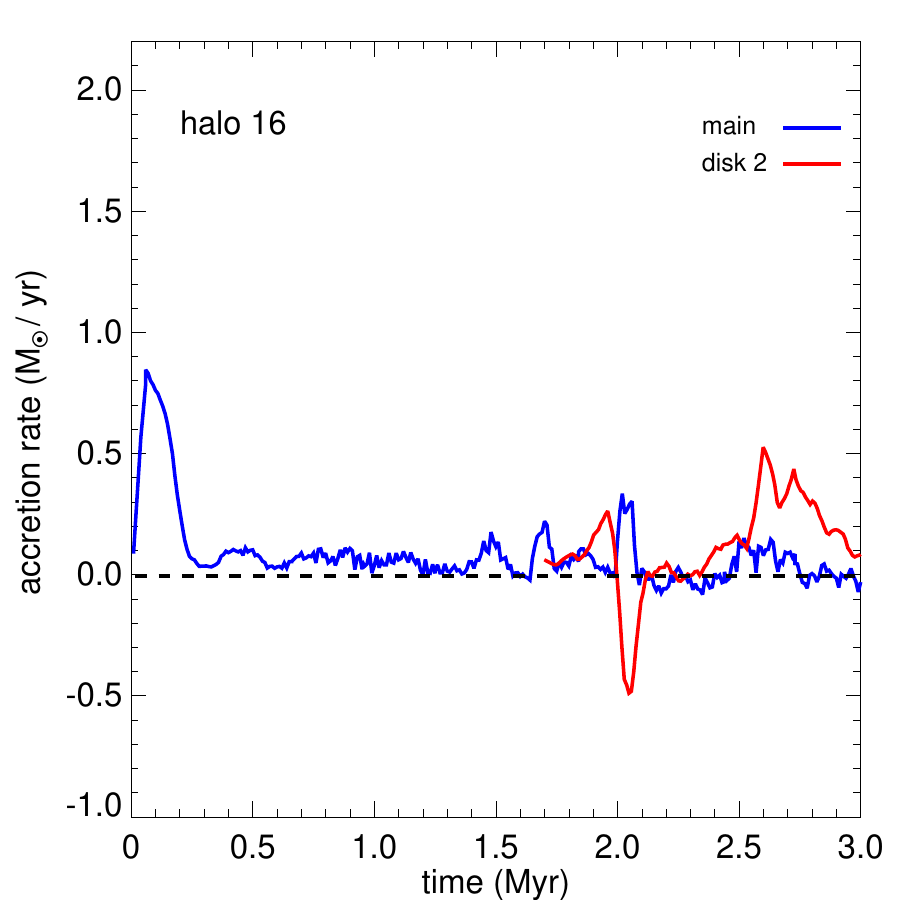,width=0.32\linewidth,clip=}  \\
\epsfig{file=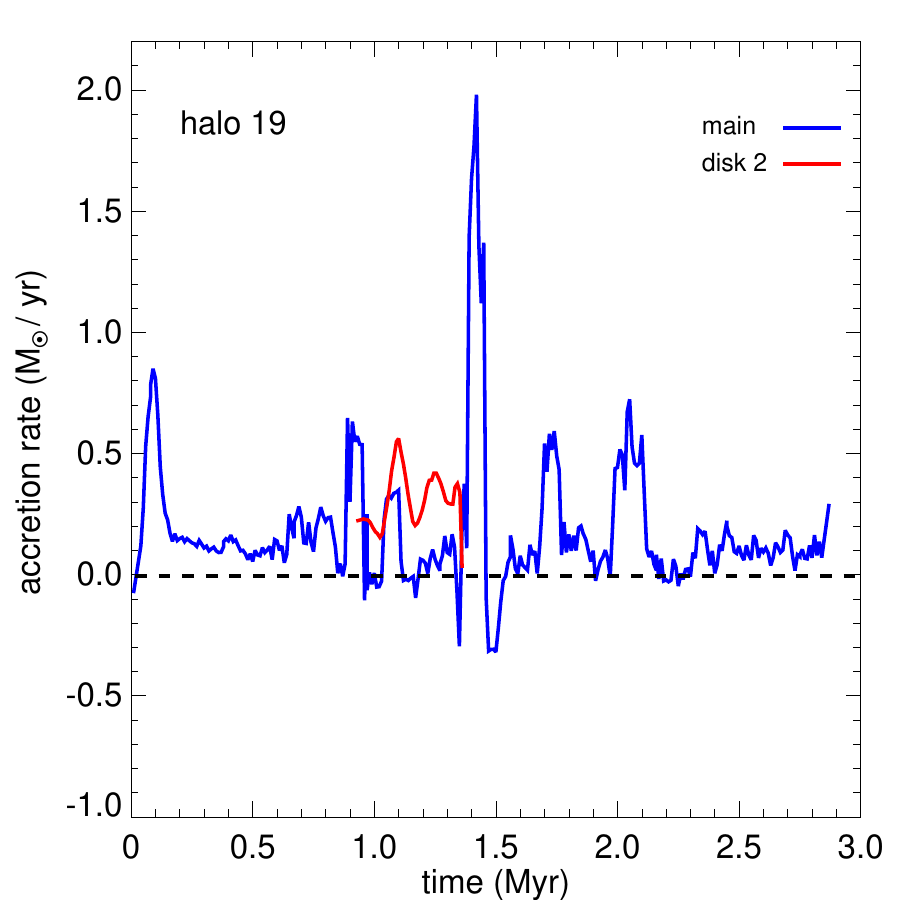,width=0.32\linewidth,clip=}  &  &   &  &
\epsfig{file=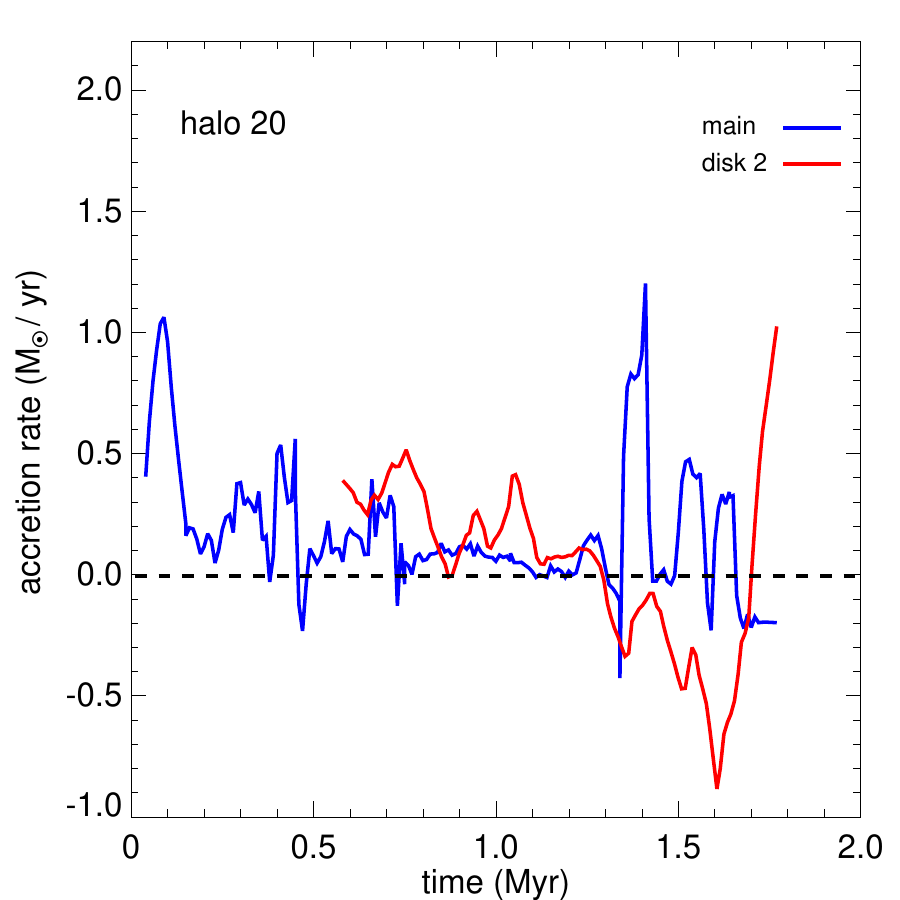,width=0.32\linewidth,clip=}  
\end{tabular}
\end{center}
\caption{Accretion rates at the centers of the main and binary disks in all eight haloes.  \label{fig:mdot}}
\end{figure*}

\begin{figure*}
\begin{center}
\begin{tabular}{cc}
\epsfig{file=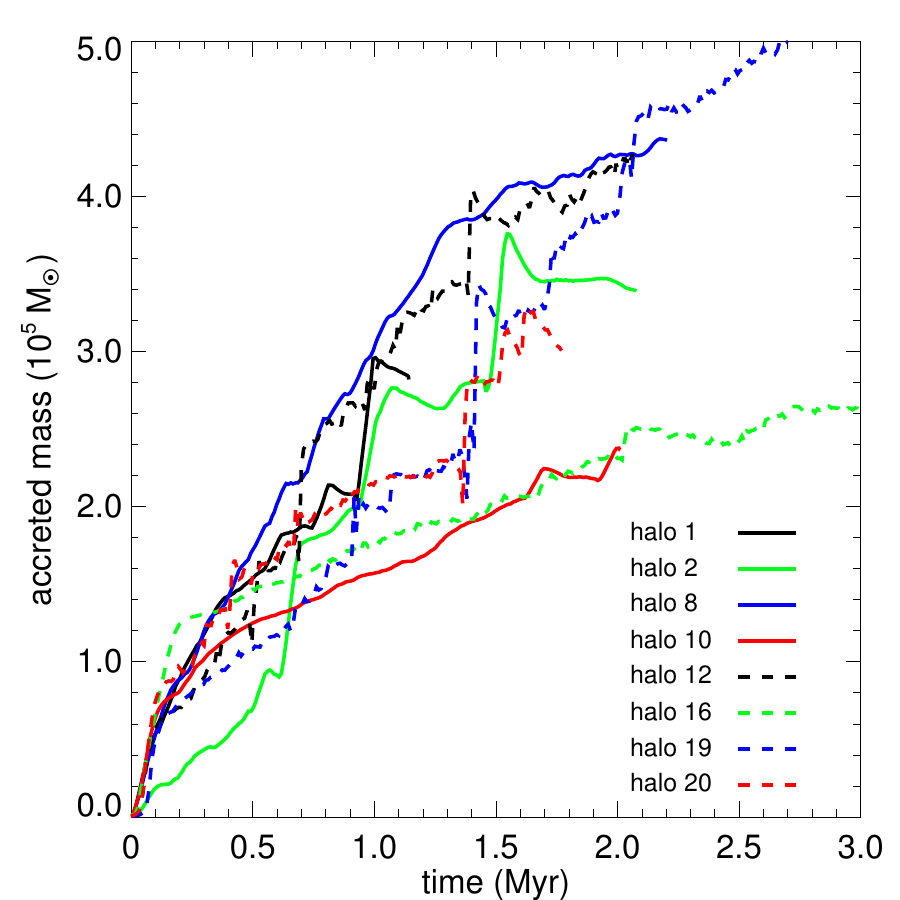,width=0.47\linewidth,clip=}  &  
\epsfig{file=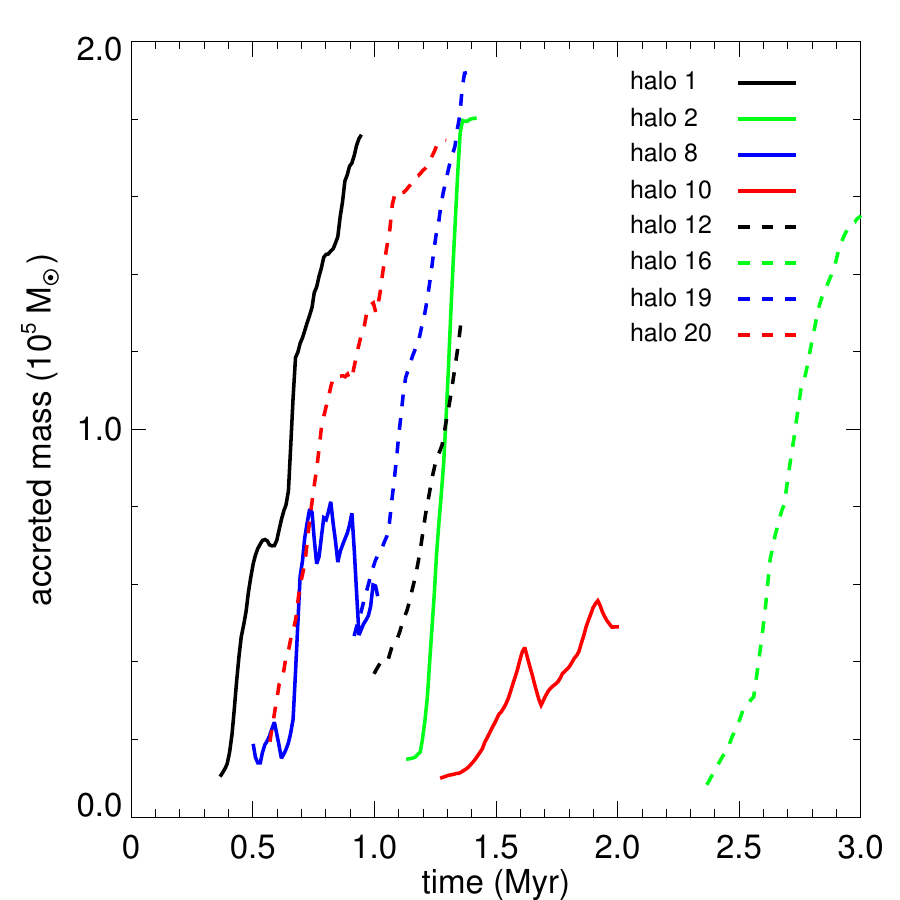,width=0.47\linewidth,clip=}  \\
\end{tabular}
\end{center}
\caption{Total mass accreted at the center of the disks over time. Left:  main disks. Right:  satellite disks.}
\label{fig:mtot}
\end{figure*}

\subsection{Accretion Rates}

We plot accretion rates for all the disks in our haloes in Figure~\ref{fig:mdot}. They are calculated by dividing the change in mass in a 0.135 pc sphere centered on the densest cell at the center of the disk over 10 kyr intervals by 10 kyr, the time between data dumps. We use this radius to ensure that the tally sphere is always resolved by at least 10 zones while excluding the spiral arms of the disk, which could produce spurious contributions to the infall rates. We also found that the accretion rates converged at this radius as we decreased the radius.  The initial time $t=0.0$ is when the densest cell first reaches the maximum refinement level and catastrophic collapse has begun.

\subsubsection{Main Disk}

In each case there is an initial jump in accretion of 0.3 - 1 \Ms\ yr$^{-1}$ for 200 - 300 kyr that is due to the formation of the dense core of the disk, like the one shown at 51.5 kyr in halo 12 in Figure~\ref{fig:evol}. Rates then fall as conservation of angular momentum of the gas increases its rotational velocity and it flattens into a self-gravitating, rotationally-supported disk, which hinders further collapse. Over the next few hundred kyr, accretion can be relatively smooth as in haloes 10 and 16, highly turbulent as in haloes 1, 8 and 19, or clumpy as in haloes 2, 12 and 20 because of fragmentation of the disk and collision of the fragments with its center. The larger jumps in accretion rate at 300 - 600 kyr are from turbulence (200 - 400 kyr in halo 8), mass exchange with a smaller, satellite disk (500 - 600 kyr in halo 1), or clumps crashing into its center (500 - 600 kyr in halo 12). The largest peaks, like those at 1.0 Myr in halo 1 and 1.4 Myr in halo 19, are from the collision of a satellite disk with the main disk.  

After the initial jump in accretion due to the formation of the dense core of the disk, the rates in some of the disks fall to lower values and flatten out as gas flows circling the center of the disk become Keplerian.  This effect is most prominent in the disks in haloes 10 and 16.  Other disks are too turbulent for the flows to ever become Keplerian and tidal interactions with satellite disks or collisions of fragments with the center of the disk disrupt circularised flows at later times, even in disks 10 and 16.

At times, the accretion rate can become negative if turbulence drives gas out of the center of the disk (and thus out of the tally sphere) or during mass exchange with another disk in a close encounter. This process occurs on much larger scales than infall onto the star itself, which would not be expected to lose mass during these episodes. During such times accretion onto the star would at most be temporarily halted. Average accretion rates after the initial surge vary from $\sim$ 0.1 - 0.5 \Ms\ yr$^{-1}$ but can peak at up to 2 \Ms\ yr$^{-1}$ for up to 100 kyr. Our rates are consistent with those reported at earlier times in previous studies \citep[$\gtrsim$ 0.1 \Ms\ yr$^{-1}$; e.g.,][]{latif15b}. Note that these average infall rates would produce SMSs that collapse via the general-relativistic instability rather than depletion of their hydrogen fuel \citep[$\gtrsim$ 0.04 \Ms\ yr$^{-1}$;][]{tyr17}.

We show the mass accumulated at the center of the main disk in each halo in the left panel of Figure~\ref{fig:mtot}. The final masses at the end of the runs vary from 2.4 $\times$ 10$^5$ \Ms\ to over 5 $\times$ 10$^5$ \Ms. The occasional dips in mass correspond to the episodes of negative accretion rate visible in Figure~\ref{fig:mdot}. Comparison of these masses to the final SMS masses plotted as a function of accretion rate in Figure 4 of \citet{tyr17} indicates that all the disks have been evolved for sufficient times for SMSs at their centers to have collapsed to DCBHs prior to the end of the run.

\subsubsection{Satellite Disks}

Accretion rates for the longest-lived binary disks in our haloes are shown in red in Figure~\ref{fig:evol}. Although they only live for a fraction of the time that the main disk evolves, their infall rates can rival those of the main disk. Peaks in accretion in the satellite disks often coincide with dips in the main disk and peaks in the main disk often coincide with dips in the satellite. These correlations arise from mass exchange between the disks during close encounters, like the ones at 540 kyr and 879 kyr in halo 1 in Figure~\ref{fig:evol}. Other peaks in accretion rate are due to collisions between the satellite disk and other clumps orbiting the main disk. Both mass exchange and collisions can perturb the orbits of binary disks. Accretion in the binary companion disks ends when they collide with the center of the main disk or are destroyed by tidal forces.

The masses accumulated at the centers of the binary disks range from approximately 75,000 \Ms\ to nearly 200,000 \Ms\ and are shown in the right panel of Figure~\ref{fig:mtot}. At our resolution (and without a detailed stellar evolution model) it is not clear how much of this mass is taken up into a star at the center of the disk, but it is likely that at least some of these disks host SMSs. As noted earlier, in some cases such as halo 19 the second SMS would form early and coevolve with the star at the center of the main disk, possibly producing an SMS binary when the second disk later crashes into the first. In other cases, such as halo 16, the second SMS would likely form after the first collapses to a DCBH and an SMS - DCBH binary would form.  

\subsubsection{Angular Momentum}

\begin{figure}
\centering
\includegraphics[width=0.47\textwidth]{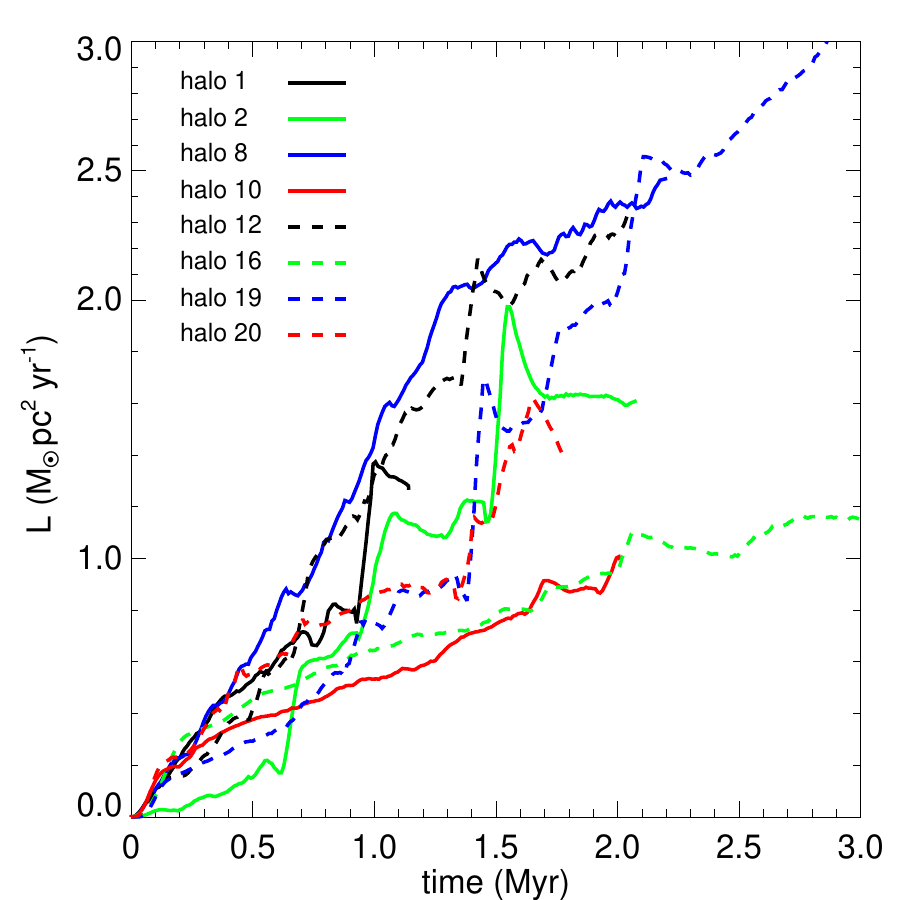}
\caption{The evolution of angular momentum in the central 0.136 pc of the disks over time.}
\label{fig:Ldisk}
\end{figure}

The evolution of the angular momentum, $L$, in the central 0.136 pc of the main disk in all eight haloes is shown in Figure~\ref{fig:Ldisk}. It mostly tracks the enclosed mass (compare with Figure~\ref{fig:mtot}), implying that no large external torques are exerted on the disk that could change the angular momentum without mass transfer. The large jumps in $L$ coincide with the collisions of massive clumps or satellite disk with the main disk, which also coincide with the large jumps in accretion rates in the disks. The subsequent drops in $L$ are due to the ejection of tidal streams after the collision that transport $L$ out of the center of the disk. There is no apparent correlation between halo spin at the onset of atomic cooling and large jumps in $L$ because they are present in halo 19, which has the lowest of the spin parameters, and halo 12, which has the second-highest spin. Likewise, there is no obvious connection between previous mergers and sudden increases in $L$ because haloes 1 and 16 both exhibit mergers during assembly but not jumps in $L$.

Note that the angular momenta shown in Figure~\ref{fig:Ldisk} are likely not taken up by the SMS at the center of the disk, which we do not resolve in our models.  Previous studies going to higher resolution for shorter times report that bars-within-bars instabilities arise on scales that are intermediate to those of the star and center of the disk that rapidly transport angular momentum outward \citep[e.g.,][]{wta08}. If such mechamisms did not arise SMSs could not form at the center of the disk because they can only accrete at most a few percent of the Keplerian angular momentum without being destroyed by the $\Omega \Gamma$ radiative instability \citep{hle18a}.

\subsection{Disk Stability}

The gravitational stability of disks is often parametrised by the radially-dependent Toomre Q parameter \citep{toomre64},
\begin{equation*}
Q(r) = \frac{\sqrt{c^2_\mathrm{s} + v^2_\mathrm{turb}} \; \kappa}{\pi G \sigma},
\end{equation*}
where $c_\mathrm{s}$ is the sound speed, $v_\mathrm{turb}$ is the turbulent speed, $\sigma$ is the surface density, and $\kappa$ is the epicyclic frequency \citep{bt87,oh02}, 
\begin{equation*}
\kappa = \sqrt{\left(2 \frac{V_\phi}{r}\right) \left(\frac{v_\phi}{r} + \frac{dv_\phi}{dr}\right)}.
\end{equation*}
If the rotation of the disk is Keplerian, $\kappa$ reduces to the rotational frequency. The Q parameter represents the ability of restorative forces in the disk (rotation and pressure gradients) to counteract gravitational perturbations. If Q falls below unity, perturbations in the disk grow rather than dampen, and it is prone to fragmentation at that radius.

\begin{figure} 
\begin{tabular}{c}
\epsfig{file=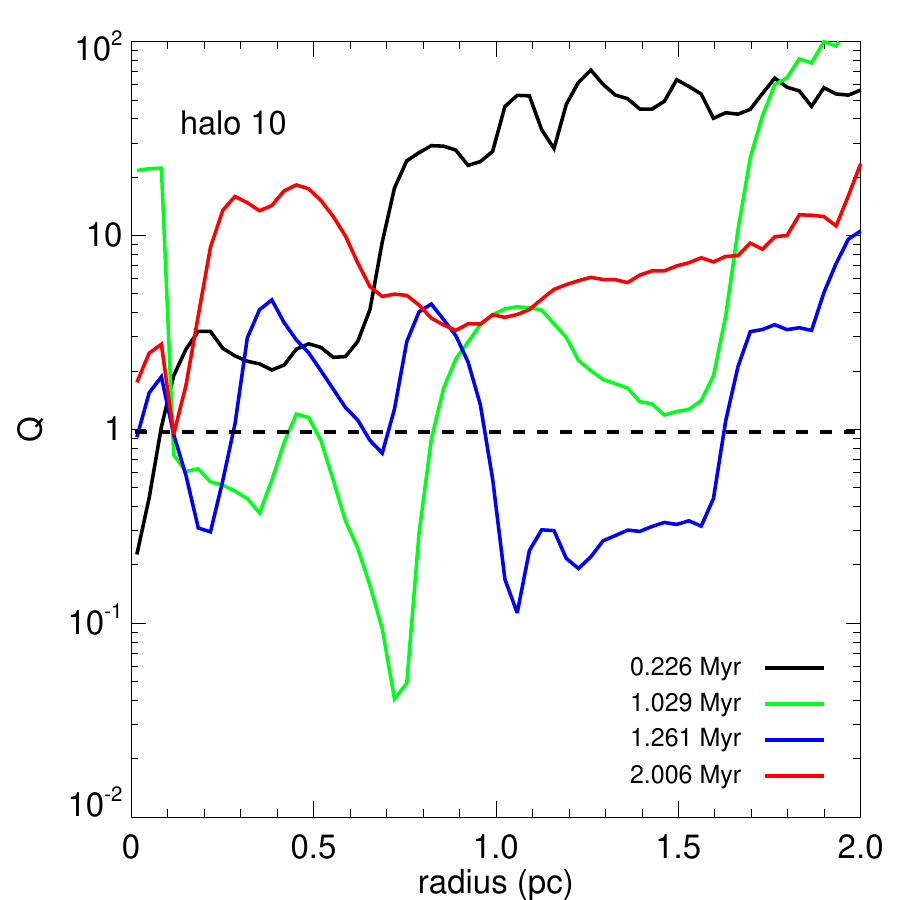,width=0.876\linewidth,clip=}  \\
\epsfig{file=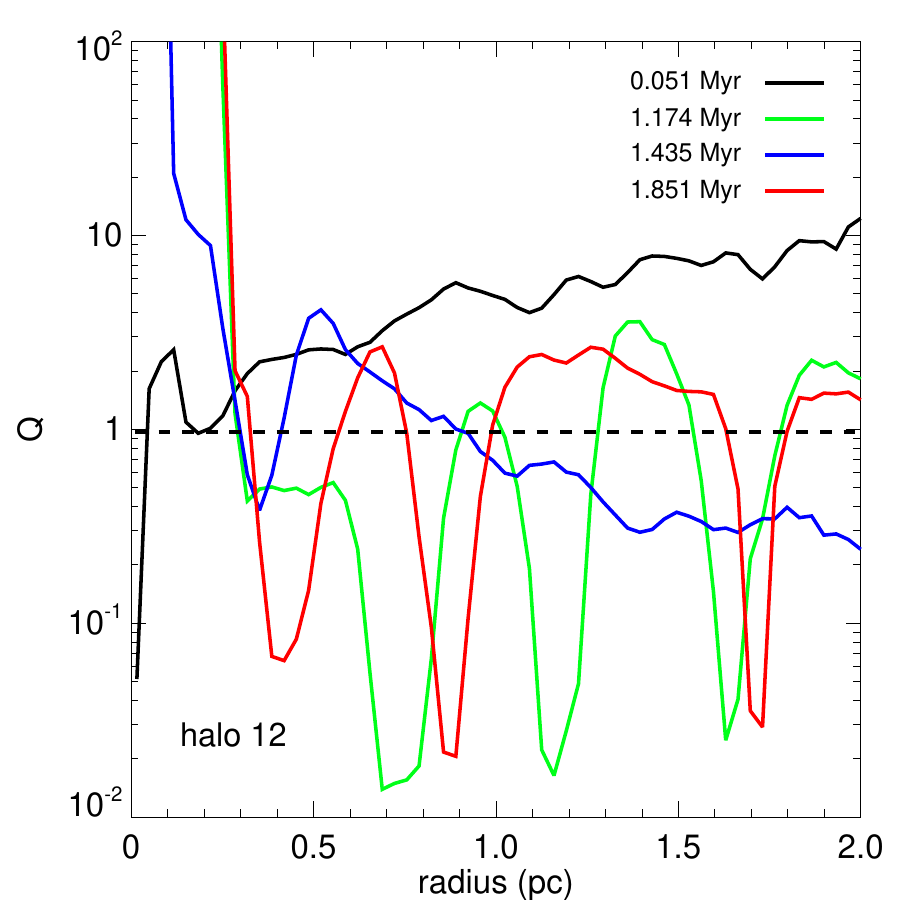,width=0.876\linewidth,clip=}  \\
\epsfig{file=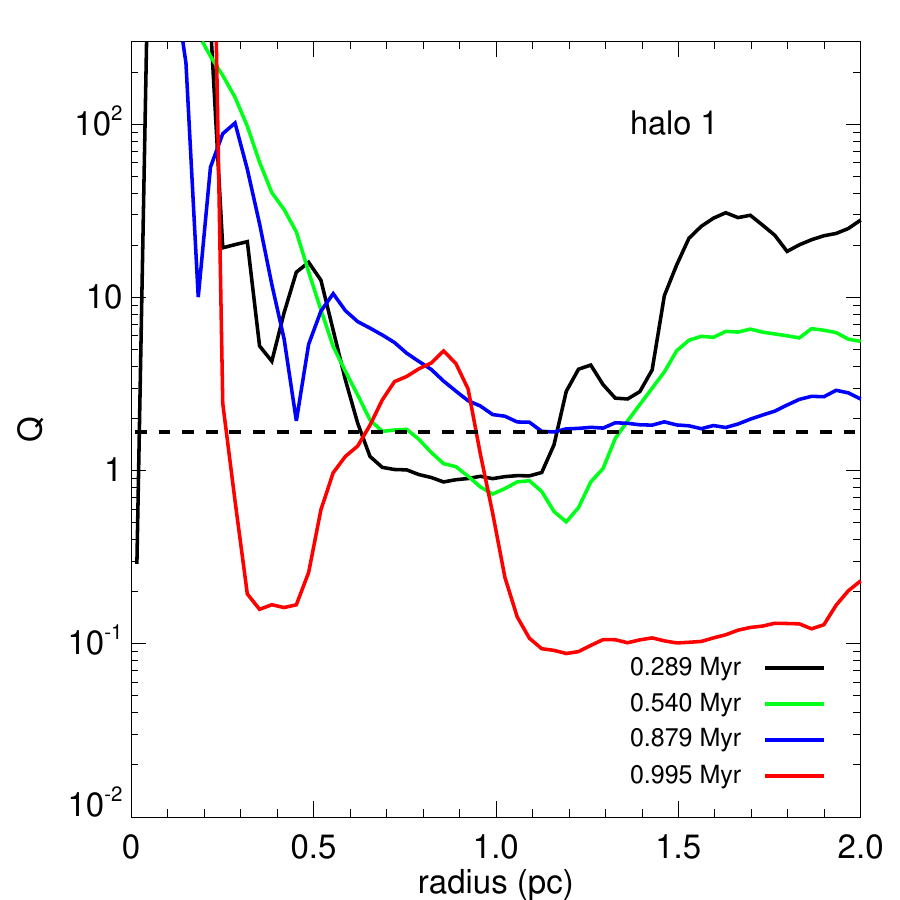,width=0.876\linewidth,clip=}  
\end{tabular}
\caption{Toomre Q parameter. Top:  halo 10 at 0.226 Myr, 1.029 Myr, 1.261 Myr and 2.006 Myr. Center:  halo 12 at 0.051 Myr, 1.174 Myr, 1.435 Myr and 1.851 Myr. Bottom:  halo 1 at 0.289 Myr, 0.540 Myr, 0.879 Myr and 0.995 Myr.  All these times correspond to those in Figure~\ref{fig:evol}.
\label{fig:toomre}}
\end{figure}

\begin{figure*}
\begin{center}
\begin{tabular}{cc}
\epsfig{file=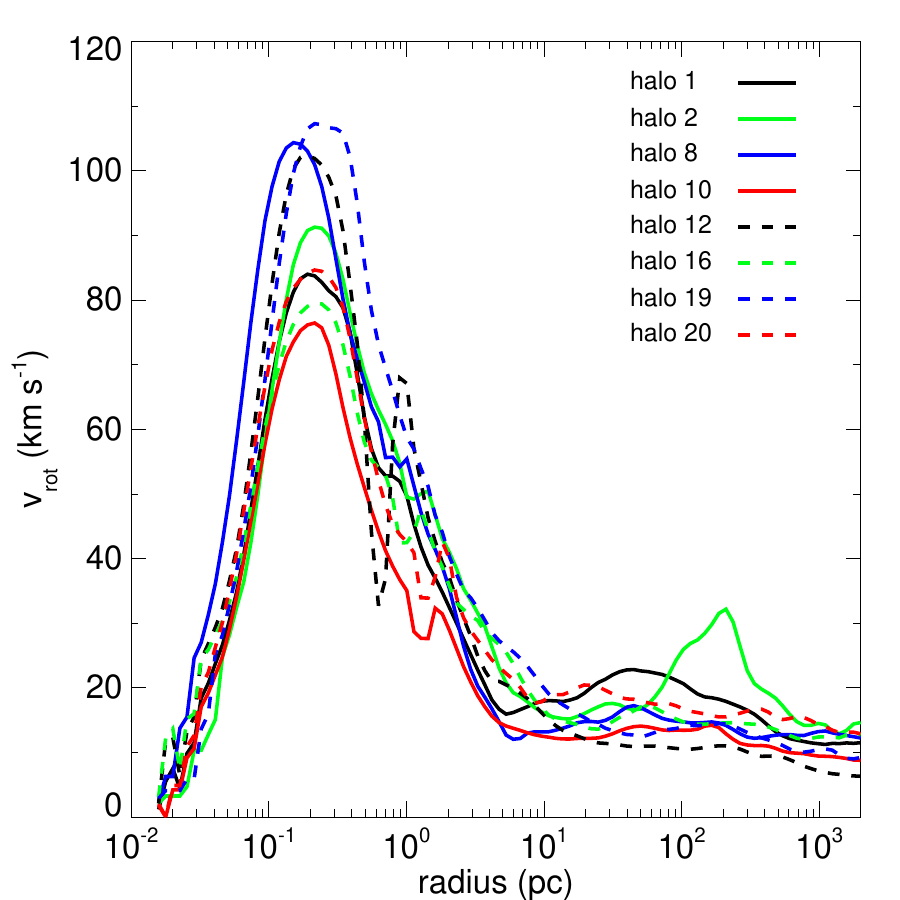,width=0.47\linewidth,clip=}  &  
\epsfig{file=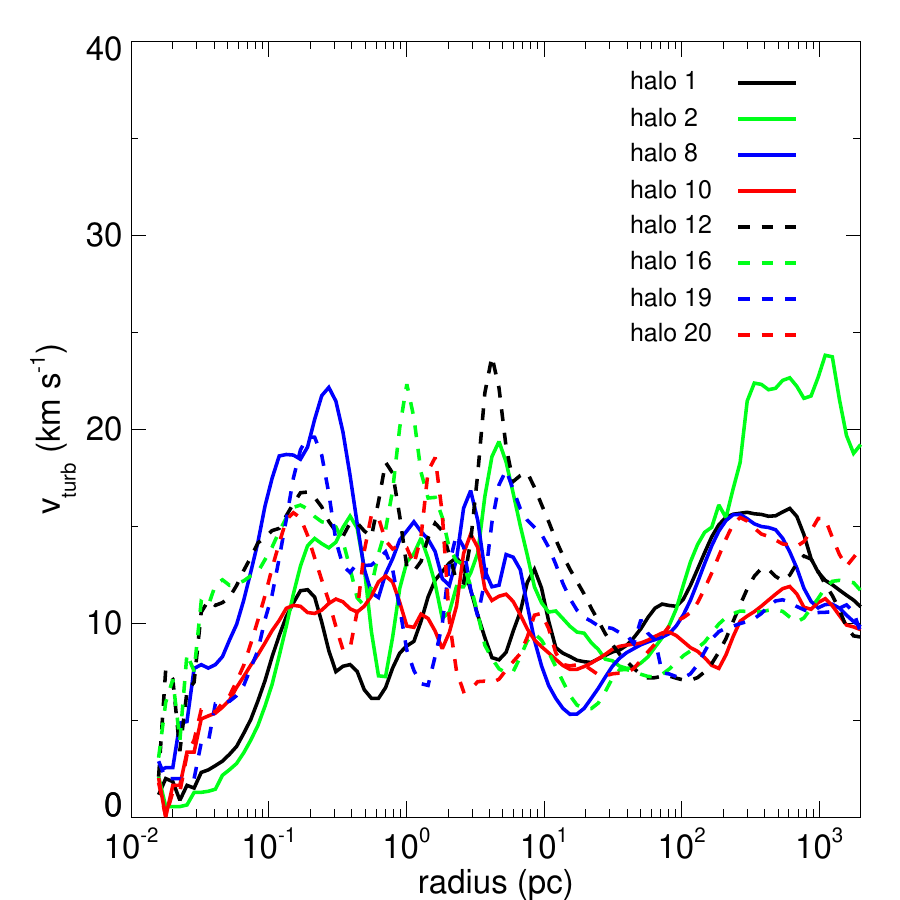,width=0.47\linewidth,clip=}  \\
\end{tabular}
\end{center}
\caption{Left: final rotational velocities as a function of radius. Right: final turbulent velocities.}
\label{fig:vrot}
\end{figure*}

Radial profiles of the Toomre parameter for the disks in haloes 10, 12, and 1 are shown in Figure~\ref{fig:toomre} for the times in Figure~\ref{fig:evol}. At 0.226 kyr in halo 10, $Q$ is greater than 1 at all radii and, as shown in Figure~\ref{fig:evol}, the disk is stable.  At 1.029 Myr $Q$ dips below 1 at $\sim$ 0.25 pc and 0.75 pc and the disk becomes prone to gravitiational instabilities that distort the spiral arms near their base and create turbulence but do not break them up or produce long-lived clumps at this time.  However, the instabilities persist and later cause the disk to fragment at 1.14 Myr and produce the clumps visible at 1.261 Myr.  $Q$ falls below 1 over a greater range of radii by this time because of the growth of the disk in mass and surface area but it again exceeds 1 everywhere by 2.006 Myr, when the fragments have coalesced into a satellite disk whose orbit around the main disk has stabilised it against fragmentation.

In contrast to halo 10, the disk in halo 12 exhibits multiple episodes of fragmentation. $Q$ falls below 1 at several radii at 1.174 Myr and again at 1.851 Myr in halo 12 and clumps are visible at those radii in Figure~\ref{fig:evol}.  Outward transport of angular momentum due to the destruction of clumps by tidal torques from the disk stabilise the disk somewhat at 1.435 Myr, when tidal tails are visible instead of fragments, but $Q$ has again fallen below 1 at 0.4 pc and beyond 0.9 pc at this time, portending the next round of fragmentation at 1.851 Myr.  As in halo 10, as the disk in halo 12 grows in mass more of it becomes subject to gravitational instabilities. 

The Toomre parameter falls below 1 by 0.289 Myr at radii of 0.5 pc - 1.1 pc in halo 1, when the disk has indeed produced the two massive clumps in Figure~\ref{fig:evol}.  The disk continues to be unstable at these radii at 0.540 Myr but then becomes stable at 1.261 Myr after the clumps have merged to produce a satellite disk who mass rivals that of the main disk.  Their mutual orbit suppresses fragmentation in both disks.  The satellite then merges with the main disk by 0.995 Myr, which ejects the tidal tails in Figure~\ref{fig:evol}, and it again becomes prone to fragmentation as $Q$ again falls below 1 at $\sim$ 0.4 pc and beyond 1.1 pc.

\subsection{Disk Rotation / Turbulence}

Rotational velocities for the main disks in all eight haloes at the end of the simulations are shown in the left panel of Figure~\ref{fig:vrot}. After multiple dynamical times the disks have all settled into rotationally supported structures with mostly Keplerian velocities. Rotational velocities are greatest at 0.1 - 0.2 pc and reach 75 - 110 km s$^{-1}$. The highest rotation speeds occur in the disks with the largest accretion rates, such as halo 8, while disks exhibiting the the lowest rotation rates have the lowest accretion rates, such as haloes 10 and 16. These velocities are consistent with those in \citet{rh09b}, who report rotation rates of up to 60 km s$^{-1}$, with higher velocities corresponding to more massive disks. Their velocities are somewhat lower than ours because their disks are evolved for much shorter times, when they have only grown to 0.3 - 0.6 pc in radius. 

The peaks and dips in velocity visible in some of the haloes at the end of the run are due to satellite disks that are orbiting the main disk at faster or slower rates. The absence of these features in the central 0.3 - 0.5 pc, along with our Toomre profiles, indicate that fragmentation does not occur near the centers of the disks at later times. The bump at $\sim$ 200 pc in halo 2 is due to an interloper that later merges with the halo. The widths of the peaks in velocity mark the outer boundaries of the disks, as gas at larger radii falling onto the disks has much lower rotation rates.

We calculate turbulent velocities in the disks by subtracting the infall and rotational velocities from the absolute velocities in quadrature: 
\begin{equation*}
v_{\mathrm{RMS}}^2 = v_{\mathrm{abs}}^2 - v_{\phi}^2 - v_r^2. 
\end{equation*}
These profiles are shown for all eight disks at the end of the simulation in the right panel of Figure~\ref{fig:vrot}. They peak at slightly larger radii than the rotational velocities, 0.2 - 0.3 pc, as gas from the spiral arms crashes into the center of the disk and the kinetic energy of bulk inflow builds up turbulent cascades and the gas becomes chaotic. Turbulent motions transport angular momentum outward through shocks that dissipate kinetic energy and they help support the disks against further collapse.

\subsection{Disk Evolution vs. Halo Assembly History}

Although we find a variety of disk evolution ranging from relatively stable disks to violently fragmenting ones, for the most part there is no apparent correlation between their evolution and the assembly histories or spin parameters of our eight haloes. For example, we find stable disks in haloes with high and low spin parameters and with as few as one and as many as three major mergers in their assembly histories (haloes 10 and 16, respectively). Likewise, disks in haloes 1 and 20 both fragment at early times but halo 1 has twice the spin parameter of halo 20. Furthermore, there is no obvious connection between longevity of binary disks and halo spin. Halo 1 has a much longer-lived satellite disk than halo 2 but a similar spin parameter. Baryons become gravitationally decoupled from DM in atomically-cooled haloes on scales of 5 - 10 pc so the evolution of the disk is governed by local gas flows and turbulence rather than DM dynamics on larger scales.

DM dynamics does appear to affect the evolution of the disks in haloes 10 and 16. From Figure~\ref{fig:mtot} it can be seen that they are only half as massive as those in the other haloes and have correspondingly lower central accretion rates. These haloes begin to atomically cool at $z =$ 17.2 and 16.5, just after major mergers at $z =$ 18 and 17.5 that give them the highest spin parameters of all the haloes. The gas inherits some of this angular momentum as it collapses to form the disks, which have more rotational support due to the larger centrifugal forces on the gas. The disks consequently grow more slowly and accumulate less mass at their centers.  

\section{Discussion and Conclusion}

\citet{suaz19} studied fragmentation in three primordial haloes that grew primarily by accretion, major mergers, or some mix of the two in a variety of LW backgrounds.  They found no fragmentation in the highest backgrounds, which approximated atomic cooling, but only evolved these haloes for a few hundred kyr.  Our results are consistent with their models because fragmentation generally happened after these times in our disks.  Binary and multiple SMSs have previously been shown to form in atomically-cooled disks at high redshift, but only in the special case of haloes with high spin parameters \citep{latif20a}.    Our new simulations prove that multiple SMS formation was the rule rather than the exception in atomically-cooled haloes because it occurred over their full range of spin parameters and for a variety of assembly histories.  Fragmentation in the vicinity of the star on AU scales has also recently been used to invoke the formation of multiple massive stars instead of a single SMS \citep[e.g.,][]{rd18}. However, they could only follow the evolution of the clumps for at most a few hundred kyr, so it is not clear if they grow to large enough masses to become stars or are just taken up into the central object, as are most of the fragments in our models on larger scales. Regardless of the final fates of fragments on small scales, our simulations show that binaries or small multiples will eventually form in the halo on larger scales.

Because DCBHs in our models typically end up in close proximity to each other after formation (0.1 - 0.2 pc) and in high ambient densities ($\gtrsim$ 10$^7$ cm$^{-3}$), radiative drag forces would rapidly decay their orbits and produce mergers whose GW emission could be detected by future observatories such as LISA \citep{til18a} in addition to NIR or radio from the BHs themselves \citep{pac15,pac17,nat17,bar18,wet20a,wet20b,wet21a,vik22a} or their progenitor stars \citep{sur18a,sur19a}.  If a SMS assumes a highly elliptical orbit around a DCBH companion it could be destroyed in a tidal disruption event (TDE).  Luminosities for TDEs of 10 - 40 \Ms\ Pop III stars with DCBHs have been calculated \citep{ki16} but not those for an SMS, which could produce an extended afterglow because of its large mass, so it is not known if such events would produce enough NIR flux to be detected today.  Both types of TDEs may be possible in a given halo because X-rays from the DCBH are known to trigger lower-mass Pop III star formation in its vicinity at later times \citep{mba03,aycin14,aycin20}.

Using the accretion rates in Figure~\ref{fig:mdot} in the Kepler stellar evolution code \citep{kep1,kep2}, we have determined that the SMSs in our disks would all evolve as cool, red supergiants with little ionizing UV that could affect their growth rates \citep{tyr22a,tyr21a}.  Consequently, radiation transport was never required in our Enzo models to obtain their true final masses.  We do not include magnetic fields in our simulations, which are thought to suppress fragmentation on small scales and result in fewer, more massive objects \citep{turk12,latif13b,latif14b,sharda20}.  However, magnetic dynamos typically affect gas flows on much smaller spatial scales than those of clump formation in our models so their absence in our simulations probably did not affect fragmentation.

\citet{tyr22a} have now shown that SMS - SMS, SMS - DCBH, and DCBH - DCBH binaries could form in our haloes.  Supermassive X-ray binaries with unique spectra could result from SMS - DCBH systems if they are in close enough proximity.  Because we do not resolve flows close to the surface of the star, our accretion rates should be taken to be upper limits. Nevertheless, as noted in the Introduction, numerical simulations that follow the collapse of atomically-cooled flows down to protostellar scales for short times find that accretion proceeds at rates similar to ours even down to these radii because of efficient angular momentum transport by bar instabilities.

We only considered isothermal collapse, in which no H$_2$ is present.  In reality, it is difficult to destroy all H$_2$ in the cores of massive haloes because of self shielding, even with very high LW backgrounds, and low to intermediate LW backgrounds were far more common in the early Universe.  Studies have shown that even small amounts of H$_2$ can dramatically reduce central infall rates that only produce 3000 - 30,000 \Ms\ stars, not 100,000 \Ms\ stars \citep[see][]{latif21a}.  There may thus have been two or three tiers of DCBH mass that led to populations of less massive quasars at high redshifts, not just the 10$^9$ \Ms\ SMBHs at $z >$ 6.  H$_2$ cooling also promotes fragmentation of the gas that could create  even less massive stars.  Given that primordial haloes fragment even in the absence of H$_2$ cooling, it was likely a common occurence in all primordial haloes.  We are investigating long-term disk evolution in a grid of LW backgrounds in all eight of our haloes now.

\section{Acknowledgements}

The authors thank the anonymous referee, whose critique improved the quality of this paper, and Britton Smith for valuable advice on the use of ytree. S. J. P. was supported by STFC grant ST/N504245/1. D. J. W. was supported by the Ida Pfeiffer Professorship at the Institute of Astrophysics at the University of Vienna and by STFC New Applicant Grant ST/P000509/1. M. A. L. thanks the UAEU for funding via UPAR grant No. 31S390 and startup grant No 31S372. All numerical simulations were done on the Sciama High Performance Computing cluster, which is supported by ICG and the University of Portsmouth.

\section{Data Availability Statement}

The data in this study will be made available upon request to the corresponding author.

\bibliographystyle{mnras}
\bibliography{refs}

\begin{thebibliography}{}
\makeatletter
\relax
\def\mn@urlcharsother{\let\do\@makeother \do\$\do\&\do\#\do\^\do\_\do\%\do\~}
\def\mn@doi{\begingroup\mn@urlcharsother \@ifnextchar [ {\mn@doi@}
  {\mn@doi@[]}}
\def\mn@doi@[#1]#2{\def\@tempa{#1}\ifx\@tempa\@empty \href
  {http://dx.doi.org/#2} {doi:#2}\else \href {http://dx.doi.org/#2} {#1}\fi
  \endgroup}
\def\mn@eprint#1#2{\mn@eprint@#1:#2::\@nil}
\def\mn@eprint@arXiv#1{\href {http://arxiv.org/abs/#1} {{\tt arXiv:#1}}}
\def\mn@eprint@dblp#1{\href {http://dblp.uni-trier.de/rec/bibtex/#1.xml}
  {dblp:#1}}
\def\mn@eprint@#1:#2:#3:#4\@nil{\def\@tempa {#1}\def\@tempb {#2}\def\@tempc
  {#3}\ifx \@tempc \@empty \let \@tempc \@tempb \let \@tempb \@tempa \fi \ifx
  \@tempb \@empty \def\@tempb {arXiv}\fi \@ifundefined
  {mn@eprint@\@tempb}{\@tempb:\@tempc}{\expandafter \expandafter \csname
  mn@eprint@\@tempb\endcsname \expandafter{\@tempc}}}

\bibitem[\protect\citeauthoryear{{Agarwal}, {Khochfar}, {Johnson}, {Neistein},
  {Dalla Vecchia}  \& {Livio}}{{Agarwal} et~al.}{2012}]{agarw12}
{Agarwal} B.,  {Khochfar} S.,  {Johnson} J.~L.,  {Neistein} E.,  {Dalla
  Vecchia} C.,   {Livio} M.,  2012, \mn@doi [\mnras]
  {10.1111/j.1365-2966.2012.21651.x}, \href
  {http://adsabs.harvard.edu/abs/2012MNRAS.425.2854A} {425, 2854}

\bibitem[\protect\citeauthoryear{{Alvarez}, {Wise}  \& {Abel}}{{Alvarez}
  et~al.}{2009}]{awa09}
{Alvarez} M.~A.,  {Wise} J.~H.,   {Abel} T.,  2009, \mn@doi [\apjl]
  {10.1088/0004-637X/701/2/L133}, \href
  {http://adsabs.harvard.edu/abs/2009ApJ...701L.133A} {701, L133}

\bibitem[\protect\citeauthoryear{{Anninos}, {Zhang}, {Abel}  \&
  {Norman}}{{Anninos} et~al.}{1997}]{anet97}
{Anninos} P.,  {Zhang} Y.,  {Abel} T.,   {Norman} M.~L.,  1997, \mn@doi [New
  Astronomy] {10.1016/S1384-1076(97)00009-2}, \href
  {http://adsabs.harvard.edu/abs/1997NewA....2..209A} {2, 209}

\bibitem[\protect\citeauthoryear{{Ardaneh}, {Luo}, {Shlosman}, {Nagamine},
  {Wise}  \& {Begelman}}{{Ardaneh} et~al.}{2018}]{ard18}
{Ardaneh} K.,  {Luo} Y.,  {Shlosman} I.,  {Nagamine} K.,  {Wise} J.~H.,
  {Begelman} M.~C.,  2018, \mn@doi [\mnras] {10.1093/mnras/sty1657}, \href
  {http://adsabs.harvard.edu/abs/2018MNRAS.479.2277A} {479, 2277}

\bibitem[\protect\citeauthoryear{{Aykutalp}, {Wise}, {Spaans}  \&
  {Meijerink}}{{Aykutalp} et~al.}{2014}]{aycin14}
{Aykutalp} A.,  {Wise} J.~H.,  {Spaans} M.,   {Meijerink} R.,  2014, \mn@doi
  [\apj] {10.1088/0004-637X/797/2/139}, \href
  {http://adsabs.harvard.edu/abs/2014ApJ...797..139A} {797, 139}

\bibitem[\protect\citeauthoryear{{Aykutalp}, {Barrow}, {Wise}  \&
  {Johnson}}{{Aykutalp} et~al.}{2020}]{aycin20}
{Aykutalp} A.,  {Barrow} K. S.~S.,  {Wise} J.~H.,   {Johnson} J.~L.,  2020,
  \mn@doi [\apjl] {10.3847/2041-8213/aba62f}, \href
  {https://ui.adsabs.harvard.edu/abs/2020ApJ...898L..53A} {898, L53}

\bibitem[\protect\citeauthoryear{{Ba{\~n}ados} et~al.,}{{Ba{\~n}ados}
  et~al.}{2018}]{ban18}
{Ba{\~n}ados} E.,  et~al., 2018, \mn@doi [\nat] {10.1038/nature25180}, \href
  {http://adsabs.harvard.edu/abs/2018Natur.553..473B} {553, 473}

\bibitem[\protect\citeauthoryear{{Barrow}, {Aykutalp}  \& {Wise}}{{Barrow}
  et~al.}{2018}]{bar18}
{Barrow} K. S.~S.,  {Aykutalp} A.,   {Wise} J.~H.,  2018, \mn@doi [Nature
  Astronomy] {10.1038/s41550-018-0569-y}, \href
  {https://ui.adsabs.harvard.edu/abs/2018NatAs...2..987B} {2, 987}

\bibitem[\protect\citeauthoryear{{Becerra}, {Greif}, {Springel}  \&
  {Hernquist}}{{Becerra} et~al.}{2015}]{bec15}
{Becerra} F.,  {Greif} T.~H.,  {Springel} V.,   {Hernquist} L.~E.,  2015,
  \mn@doi [\mnras] {10.1093/mnras/stu2284}, \href
  {http://adsabs.harvard.edu/abs/2015MNRAS.446.2380B} {446, 2380}

\bibitem[\protect\citeauthoryear{{Becerra}, {Marinacci}, {Bromm}  \&
  {Hernquist}}{{Becerra} et~al.}{2018}]{bec18}
{Becerra} F.,  {Marinacci} F.,  {Bromm} V.,   {Hernquist} L.~E.,  2018, \mn@doi
  [\mnras] {10.1093/mnras/sty2210}, \href
  {http://adsabs.harvard.edu/abs/2018MNRAS.480.5029B} {480, 5029}

\bibitem[\protect\citeauthoryear{{Behroozi}, {Wechsler}  \& {Wu}}{{Behroozi}
  et~al.}{2013}]{rockstar}
{Behroozi} P.~S.,  {Wechsler} R.~H.,   {Wu} H.-Y.,  2013, \mn@doi [\apj]
  {10.1088/0004-637X/762/2/109}, \href
  {https://ui.adsabs.harvard.edu/abs/2013ApJ...762..109B} {762, 109}

\bibitem[\protect\citeauthoryear{{Binney} \& {Tremaine}}{{Binney} \&
  {Tremaine}}{1987}]{bt87}
{Binney} J.,  {Tremaine} S.,  1987, {Galactic dynamics}

\bibitem[\protect\citeauthoryear{{Boekholt}, {Schleicher}, {Fellhauer},
  {Klessen}, {Reinoso}, {Stutz}  \& {Haemmerl{\'e}}}{{Boekholt}
  et~al.}{2018}]{boek18}
{Boekholt} T.~C.~N.,  {Schleicher} D.~R.~G.,  {Fellhauer} M.,  {Klessen} R.~S.,
   {Reinoso} B.,  {Stutz} A.~M.,   {Haemmerl{\'e}} L.,  2018, \mn@doi [\mnras]
  {10.1093/mnras/sty208}, \href
  {http://adsabs.harvard.edu/abs/2018MNRAS.476..366B} {476, 366}

\bibitem[\protect\citeauthoryear{{Bromm} \& {Loeb}}{{Bromm} \&
  {Loeb}}{2003}]{bl03}
{Bromm} V.,  {Loeb} A.,  2003, \mn@doi [\apj] {10.1086/377529}, \href
  {http://adsabs.harvard.edu/abs/2003ApJ...596...34B} {596, 34}

\bibitem[\protect\citeauthoryear{{Bryan} \& {Norman}}{{Bryan} \&
  {Norman}}{1997}]{bn97}
{Bryan} G.~L.,  {Norman} M.~L.,  1997, arXiv:astro-ph/9710187, \href
  {http://adsabs.harvard.edu/abs/1997astro.ph.10187B} {}

\bibitem[\protect\citeauthoryear{{Bryan}, {Norman}, {Stone}, {Cen}  \&
  {Ostriker}}{{Bryan} et~al.}{1995}]{bryan95}
{Bryan} G.~L.,  {Norman} M.~L.,  {Stone} J.~M.,  {Cen} R.,   {Ostriker} J.~P.,
  1995, \mn@doi [Computer Physics Communications]
  {10.1016/0010-4655(94)00191-4}, \href
  {http://adsabs.harvard.edu/abs/1995CoPhC..89..149B} {89, 149}

\bibitem[\protect\citeauthoryear{{Bryan} et~al.,}{{Bryan} et~al.}{2014}]{enzo}
{Bryan} G.~L.,  et~al., 2014, \mn@doi [\apjs] {10.1088/0067-0049/211/2/19},
  \href {http://adsabs.harvard.edu/abs/2014ApJS..211...19B} {211, 19}

\bibitem[\protect\citeauthoryear{{Bullock}, {Dekel}, {Kolatt}, {Kravtsov},
  {Klypin}, {Porciani}  \& {Primack}}{{Bullock} et~al.}{2001}]{bul01}
{Bullock} J.~S.,  {Dekel} A.,  {Kolatt} T.~S.,  {Kravtsov} A.~V.,  {Klypin}
  A.~A.,  {Porciani} C.,   {Primack} J.~R.,  2001, \mn@doi [\apj]
  {10.1086/321477}, \href
  {https://ui.adsabs.harvard.edu/abs/2001ApJ...555..240B} {555, 240}

\bibitem[\protect\citeauthoryear{{Chon}, {Hirano}, {Hosokawa}  \&
  {Yoshida}}{{Chon} et~al.}{2016}]{chon16}
{Chon} S.,  {Hirano} S.,  {Hosokawa} T.,   {Yoshida} N.,  2016, \mn@doi [\apj]
  {10.3847/0004-637X/832/2/134}, \href
  {https://ui.adsabs.harvard.edu/abs/2016ApJ...832..134C} {832, 134}

\bibitem[\protect\citeauthoryear{{Chon}, {Hosokawa}  \& {Yoshida}}{{Chon}
  et~al.}{2018}]{chon18}
{Chon} S.,  {Hosokawa} T.,   {Yoshida} N.,  2018, \mn@doi [\mnras]
  {10.1093/mnras/sty086}, \href
  {https://ui.adsabs.harvard.edu/abs/2018MNRAS.475.4104C} {475, 4104}

\bibitem[\protect\citeauthoryear{{Couchman}}{{Couchman}}{1991}]{couch91}
{Couchman} H.~M.~P.,  1991, \mn@doi [\apjl] {10.1086/185939}, \href
  {http://adsabs.harvard.edu/abs/1991ApJ...368L..23C} {368, L23}

\bibitem[\protect\citeauthoryear{{Devecchi} \& {Volonteri}}{{Devecchi} \&
  {Volonteri}}{2009}]{dv09}
{Devecchi} B.,  {Volonteri} M.,  2009, \mn@doi [\apj]
  {10.1088/0004-637X/694/1/302}, \href
  {http://adsabs.harvard.edu/abs/2009ApJ...694..302D} {694, 302}

\bibitem[\protect\citeauthoryear{{Di Matteo}, {Khandai}, {DeGraf}, {Feng},
  {Croft}, {Lopez}  \& {Springel}}{{Di Matteo} et~al.}{2012}]{dm12}
{Di Matteo} T.,  {Khandai} N.,  {DeGraf} C.,  {Feng} Y.,  {Croft} R.~A.~C.,
  {Lopez} J.,   {Springel} V.,  2012, \mn@doi [\apjl]
  {10.1088/2041-8205/745/2/L29}, \href
  {http://adsabs.harvard.edu/abs/2012ApJ...745L..29D} {745, L29}

\bibitem[\protect\citeauthoryear{{Di Matteo}, {Croft}, {Feng}, {Waters}  \&
  {Wilkins}}{{Di Matteo} et~al.}{2017}]{dm17}
{Di Matteo} T.,  {Croft} R. A.~C.,  {Feng} Y.,  {Waters} D.,   {Wilkins} S.,
  2017, \mn@doi [\mnras] {10.1093/mnras/stx319}, \href
  {https://ui.adsabs.harvard.edu/abs/2017MNRAS.467.4243D} {467, 4243}

\bibitem[\protect\citeauthoryear{{Dijkstra}, {Ferrara}  \&
  {Mesinger}}{{Dijkstra} et~al.}{2014}]{dfm14}
{Dijkstra} M.,  {Ferrara} A.,   {Mesinger} A.,  2014, \mn@doi [\mnras]
  {10.1093/mnras/stu1007}, \href
  {http://adsabs.harvard.edu/abs/2014MNRAS.442.2036D} {442, 2036}

\bibitem[\protect\citeauthoryear{{Efstathiou}, {Davis}, {White}  \&
  {Frenk}}{{Efstathiou} et~al.}{1985}]{efs85}
{Efstathiou} G.,  {Davis} M.,  {White} S.~D.~M.,   {Frenk} C.~S.,  1985,
  \mn@doi [\apjs] {10.1086/191003}, \href
  {http://adsabs.harvard.edu/abs/1985ApJS...57..241E} {57, 241}

\bibitem[\protect\citeauthoryear{{Fan} et~al.,}{{Fan} et~al.}{2003}]{fan03}
{Fan} X.,  et~al., 2003, \mn@doi [\aj] {10.1086/368246}, \href
  {http://adsabs.harvard.edu/abs/2003AJ....125.1649F} {125, 1649}

\bibitem[\protect\citeauthoryear{{Glover} \& {Abel}}{{Glover} \&
  {Abel}}{2008}]{ga08}
{Glover} S.~C.~O.,  {Abel} T.,  2008, \mn@doi [\mnras]
  {10.1111/j.1365-2966.2008.13224.x}, \href
  {http://adsabs.harvard.edu/abs/2008MNRAS.388.1627G} {388, 1627}

\bibitem[\protect\citeauthoryear{{Greif}, {White}, {Klessen}  \&
  {Springel}}{{Greif} et~al.}{2011}]{greif11}
{Greif} T.~H.,  {White} S.~D.~M.,  {Klessen} R.~S.,   {Springel} V.,  2011,
  \mn@doi [\apj] {10.1088/0004-637X/736/2/147}, \href
  {http://adsabs.harvard.edu/abs/2011ApJ...736..147G} {736, 147}

\bibitem[\protect\citeauthoryear{{Haemmerl{\'e}}, {Woods}, {Klessen}, {Heger}
  \& {Whalen}}{{Haemmerl{\'e}} et~al.}{2018a}]{hle18b}
{Haemmerl{\'e}} L.,  {Woods} T.~E.,  {Klessen} R.~S.,  {Heger} A.,   {Whalen}
  D.~J.,  2018a, \mn@doi [\mnras] {10.1093/mnras/stx2919}, \href
  {http://adsabs.harvard.edu/abs/2018MNRAS.474.2757H} {474, 2757}

\bibitem[\protect\citeauthoryear{{Haemmerl{\'e}}, {Woods}, {Klessen}, {Heger}
  \& {Whalen}}{{Haemmerl{\'e}} et~al.}{2018b}]{hle18a}
{Haemmerl{\'e}} L.,  {Woods} T.~E.,  {Klessen} R.~S.,  {Heger} A.,   {Whalen}
  D.~J.,  2018b, \mn@doi [\apjl] {10.3847/2041-8213/aaa462}, \href
  {http://adsabs.harvard.edu/abs/2018ApJ...853L...3H} {853, L3}

\bibitem[\protect\citeauthoryear{{Hahn} \& {Abel}}{{Hahn} \&
  {Abel}}{2011}]{hahn11}
{Hahn} O.,  {Abel} T.,  2011, \mn@doi [\mnras]
  {10.1111/j.1365-2966.2011.18820.x}, \href
  {http://adsabs.harvard.edu/abs/2011MNRAS.415.2101H} {415, 2101}

\bibitem[\protect\citeauthoryear{{Hartwig}, {Agarwal}  \& {Regan}}{{Hartwig}
  et~al.}{2018}]{til18a}
{Hartwig} T.,  {Agarwal} B.,   {Regan} J.~A.,  2018, \mn@doi [\mnras]
  {10.1093/mnrasl/sly091}, \href
  {http://adsabs.harvard.edu/abs/2018MNRAS.479L..23H} {479, L23}

\bibitem[\protect\citeauthoryear{{Herrington}, {Whalen}  \&
  {Woods}}{{Herrington} et~al.}{2023}]{herr23a}
{Herrington} N.~P.,  {Whalen} D.~J.,   {Woods} T.~E.,  2023, \mn@doi [\mnras]
  {10.1093/mnras/stad572}, \href
  {https://ui.adsabs.harvard.edu/abs/2023MNRAS.521..463H} {521, 463}

\bibitem[\protect\citeauthoryear{{Hirano}, {Hosokawa}, {Yoshida}, {Umeda},
  {Omukai}, {Chiaki}  \& {Yorke}}{{Hirano} et~al.}{2014}]{hir13}
{Hirano} S.,  {Hosokawa} T.,  {Yoshida} N.,  {Umeda} H.,  {Omukai} K.,
  {Chiaki} G.,   {Yorke} H.~W.,  2014, \mn@doi [\apj]
  {10.1088/0004-637X/781/2/60}, \href
  {http://adsabs.harvard.edu/abs/2014ApJ...781...60H} {781, 60}

\bibitem[\protect\citeauthoryear{{Hirano}, {Hosokawa}, {Yoshida}, {Omukai}  \&
  {Yorke}}{{Hirano} et~al.}{2015}]{hir15}
{Hirano} S.,  {Hosokawa} T.,  {Yoshida} N.,  {Omukai} K.,   {Yorke} H.~W.,
  2015, \mn@doi [\mnras] {10.1093/mnras/stv044}, \href
  {http://adsabs.harvard.edu/abs/2015MNRAS.448..568H} {448, 568}

\bibitem[\protect\citeauthoryear{{Hirano}, {Hosokawa}, {Yoshida}  \&
  {Kuiper}}{{Hirano} et~al.}{2017}]{hir17}
{Hirano} S.,  {Hosokawa} T.,  {Yoshida} N.,   {Kuiper} R.,  2017, \mn@doi
  [Science] {10.1126/science.aai9119}, \href
  {http://adsabs.harvard.edu/abs/2017Sci...357.1375H} {357, 1375}

\bibitem[\protect\citeauthoryear{{Hosokawa}, {Yorke}, {Inayoshi}, {Omukai}  \&
  {Yoshida}}{{Hosokawa} et~al.}{2013}]{hos13}
{Hosokawa} T.,  {Yorke} H.~W.,  {Inayoshi} K.,  {Omukai} K.,   {Yoshida} N.,
  2013, \mn@doi [\apj] {10.1088/0004-637X/778/2/178}, \href
  {http://adsabs.harvard.edu/abs/2013ApJ...778..178H} {778, 178}

\bibitem[\protect\citeauthoryear{{Inoue} \& {Yoshida}}{{Inoue} \&
  {Yoshida}}{2020}]{iy20}
{Inoue} S.,  {Yoshida} N.,  2020, \mn@doi [\mnras] {10.1093/mnrasl/slz160},
  \href {https://ui.adsabs.harvard.edu/abs/2020MNRAS.491L..24I} {491, L24}

\bibitem[\protect\citeauthoryear{{Johnson}, {Whalen}, {Li}  \&
  {Holz}}{{Johnson} et~al.}{2013}]{jet13}
{Johnson} J.~L.,  {Whalen} D.~J.,  {Li} H.,   {Holz} D.~E.,  2013, \mn@doi
  [\apj] {10.1088/0004-637X/771/2/116}, \href
  {http://adsabs.harvard.edu/abs/2013ApJ...771..116J} {771, 116}

\bibitem[\protect\citeauthoryear{{Johnson}, {Whalen}, {Agarwal}, {Paardekooper}
   \& {Khochfar}}{{Johnson} et~al.}{2014}]{jet14}
{Johnson} J.~L.,  {Whalen} D.~J.,  {Agarwal} B.,  {Paardekooper} J.-P.,
  {Khochfar} S.,  2014, \mn@doi [\mnras] {10.1093/mnras/stu1676}, \href
  {http://adsabs.harvard.edu/abs/2014MNRAS.445..686J} {445, 686}

\bibitem[\protect\citeauthoryear{{Kashiyama} \& {Inayoshi}}{{Kashiyama} \&
  {Inayoshi}}{2016}]{ki16}
{Kashiyama} K.,  {Inayoshi} K.,  2016, \mn@doi [\apj]
  {10.3847/0004-637X/826/1/80}, \href
  {http://adsabs.harvard.edu/abs/2016ApJ...826...80K} {826, 80}

\bibitem[\protect\citeauthoryear{{Kitayama}, {Yoshida}, {Susa}  \&
  {Umemura}}{{Kitayama} et~al.}{2004}]{ket04}
{Kitayama} T.,  {Yoshida} N.,  {Susa} H.,   {Umemura} M.,  2004, \mn@doi [\apj]
  {10.1086/423313}, \href {http://adsabs.harvard.edu/abs/2004ApJ...613..631K}
  {613, 631}

\bibitem[\protect\citeauthoryear{{Latif} \& {Khochfar}}{{Latif} \&
  {Khochfar}}{2020}]{latif20b}
{Latif} M.~A.,  {Khochfar} S.,  2020, \mn@doi [\mnras]
  {10.1093/mnras/staa2218}, \href
  {https://ui.adsabs.harvard.edu/abs/2020MNRAS.497.3761L} {497, 3761}

\bibitem[\protect\citeauthoryear{{Latif} \& {Volonteri}}{{Latif} \&
  {Volonteri}}{2015}]{latif15b}
{Latif} M.~A.,  {Volonteri} M.,  2015, \mn@doi [\mnras]
  {10.1093/mnras/stv1337}, \href
  {http://adsabs.harvard.edu/abs/2015MNRAS.452.1026L} {452, 1026}

\bibitem[\protect\citeauthoryear{{Latif}, {Schleicher}, {Schmidt}  \&
  {Niemeyer}}{{Latif} et~al.}{2013a}]{latif13a}
{Latif} M.~A.,  {Schleicher} D.~R.~G.,  {Schmidt} W.,   {Niemeyer} J.,  2013a,
  \mn@doi [\mnras] {10.1093/mnras/sts659}, \href
  {http://adsabs.harvard.edu/abs/2013MNRAS.430..588L} {430, 588}

\bibitem[\protect\citeauthoryear{{Latif}, {Schleicher}, {Schmidt}  \&
  {Niemeyer}}{{Latif} et~al.}{2013b}]{latif13b}
{Latif} M.~A.,  {Schleicher} D.~R.~G.,  {Schmidt} W.,   {Niemeyer} J.,  2013b,
  \mn@doi [\mnras] {10.1093/mnras/stt503}, \href
  {http://adsabs.harvard.edu/abs/2013MNRAS.432..668L} {432, 668}

\bibitem[\protect\citeauthoryear{{Latif}, {Schleicher}, {Schmidt}  \&
  {Niemeyer}}{{Latif} et~al.}{2013c}]{latif13d}
{Latif} M.~A.,  {Schleicher} D.~R.~G.,  {Schmidt} W.,   {Niemeyer} J.~C.,
  2013c, \mn@doi [\mnras] {10.1093/mnras/stt1786}, \href
  {http://adsabs.harvard.edu/abs/2013MNRAS.436.2989L} {436, 2989}

\bibitem[\protect\citeauthoryear{{Latif}, {Schleicher}  \& {Schmidt}}{{Latif}
  et~al.}{2014a}]{latif14b}
{Latif} M.~A.,  {Schleicher} D.~R.~G.,   {Schmidt} W.,  2014a, \mn@doi [\mnras]
  {10.1093/mnras/stu357}, \href
  {https://ui.adsabs.harvard.edu/abs/2014MNRAS.440.1551L} {440, 1551}

\bibitem[\protect\citeauthoryear{{Latif}, {Bovino}, {Van Borm}, {Grassi},
  {Schleicher}  \& {Spaans}}{{Latif} et~al.}{2014b}]{latif14}
{Latif} M.~A.,  {Bovino} S.,  {Van Borm} C.,  {Grassi} T.,  {Schleicher}
  D.~R.~G.,   {Spaans} M.,  2014b, \mn@doi [\mnras] {10.1093/mnras/stu1230},
  \href {http://adsabs.harvard.edu/abs/2014MNRAS.443.1979L} {443, 1979}

\bibitem[\protect\citeauthoryear{{Latif}, {Bovino}, {Grassi}, {Schleicher}  \&
  {Spaans}}{{Latif} et~al.}{2015}]{latif15a}
{Latif} M.~A.,  {Bovino} S.,  {Grassi} T.,  {Schleicher} D.~R.~G.,   {Spaans}
  M.,  2015, \mn@doi [\mnras] {10.1093/mnras/stu2244}, \href
  {http://adsabs.harvard.edu/abs/2015MNRAS.446.3163L} {446, 3163}

\bibitem[\protect\citeauthoryear{{Latif}, {Omukai}, {Habouzit}, {Schleicher}
  \& {Volonteri}}{{Latif} et~al.}{2016}]{latif16}
{Latif} M.~A.,  {Omukai} K.,  {Habouzit} M.,  {Schleicher} D.~R.~G.,
  {Volonteri} M.,  2016, \mn@doi [\apj] {10.3847/0004-637X/823/1/40}, \href
  {http://adsabs.harvard.edu/abs/2016ApJ...823...40L} {823, 40}

\bibitem[\protect\citeauthoryear{{Latif}, {Khochfar}  \& {Whalen}}{{Latif}
  et~al.}{2020}]{latif20a}
{Latif} M.~A.,  {Khochfar} S.,   {Whalen} D.,  2020, \mn@doi [\apjl]
  {10.3847/2041-8213/ab7c61}, \href
  {https://ui.adsabs.harvard.edu/abs/2020ApJ...892L...4L} {892, L4}

\bibitem[\protect\citeauthoryear{{Latif}, {Khochfar}, {Schleicher}  \&
  {Whalen}}{{Latif} et~al.}{2021}]{latif21a}
{Latif} M.~A.,  {Khochfar} S.,  {Schleicher} D.,   {Whalen} D.~J.,  2021,
  \mn@doi [\mnras] {10.1093/mnras/stab2708}, \href
  {https://ui.adsabs.harvard.edu/abs/2021MNRAS.508.1756L} {508, 1756}

\bibitem[\protect\citeauthoryear{{Latif}, {Whalen}, {Khochfar}, {Herrington}
  \& {Woods}}{{Latif} et~al.}{2022}]{latif22b}
{Latif} M.~A.,  {Whalen} D.~J.,  {Khochfar} S.,  {Herrington} N.~P.,   {Woods}
  T.~E.,  2022, \mn@doi [\nat] {10.1038/s41586-022-04813-y}, \href
  {https://ui.adsabs.harvard.edu/abs/2022Natur.607...48L} {607, 48}

\bibitem[\protect\citeauthoryear{{Luo}, {Ardaneh}, {Shlosman}, {Nagamine},
  {Wise}  \& {Begelman}}{{Luo} et~al.}{2018}]{luo18}
{Luo} Y.,  {Ardaneh} K.,  {Shlosman} I.,  {Nagamine} K.,  {Wise} J.~H.,
  {Begelman} M.~C.,  2018, \mn@doi [\mnras] {10.1093/mnras/sty362}, \href
  {http://adsabs.harvard.edu/abs/2018MNRAS.476.3523L} {476, 3523}

\bibitem[\protect\citeauthoryear{{Lupi}, {Volonteri}, {Decarli}, {Bovino},
  {Silk}  \& {Bergeron}}{{Lupi} et~al.}{2019}]{lup19}
{Lupi} A.,  {Volonteri} M.,  {Decarli} R.,  {Bovino} S.,  {Silk} J.,
  {Bergeron} J.,  2019, \mn@doi [\mnras] {10.1093/mnras/stz1959}, \href
  {https://ui.adsabs.harvard.edu/abs/2019MNRAS.488.4004L} {488, 4004}

\bibitem[\protect\citeauthoryear{{Lupi}, {Haiman}  \& {Volonteri}}{{Lupi}
  et~al.}{2021}]{lup21}
{Lupi} A.,  {Haiman} Z.,   {Volonteri} M.,  2021, \mn@doi [\mnras]
  {10.1093/mnras/stab692}, \href
  {https://ui.adsabs.harvard.edu/abs/2021MNRAS.503.5046L} {503, 5046}

\bibitem[\protect\citeauthoryear{{Machacek}, {Bryan}  \& {Abel}}{{Machacek}
  et~al.}{2001}]{met01}
{Machacek} M.~E.,  {Bryan} G.~L.,   {Abel} T.,  2001, \mn@doi [\apj]
  {10.1086/319014}, \href {http://adsabs.harvard.edu/abs/2001ApJ...548..509M}
  {548, 509}

\bibitem[\protect\citeauthoryear{{Machacek}, {Bryan}  \& {Abel}}{{Machacek}
  et~al.}{2003}]{mba03}
{Machacek} M.~E.,  {Bryan} G.~L.,   {Abel} T.,  2003, \mn@doi [\mnras]
  {10.1046/j.1365-8711.2003.06054.x}, \href
  {http://adsabs.harvard.edu/abs/2003MNRAS.338..273M} {338, 273}

\bibitem[\protect\citeauthoryear{{Maio}, {Borgani}, {Ciardi}  \&
  {Petkova}}{{Maio} et~al.}{2019}]{maio19}
{Maio} U.,  {Borgani} S.,  {Ciardi} B.,   {Petkova} M.,  2019, \mn@doi [\pasa]
  {10.1017/pasa.2019.10}, \href
  {https://ui.adsabs.harvard.edu/abs/2019PASA...36...20M} {36, e020}

\bibitem[\protect\citeauthoryear{{Matsuoka} et~al.,}{{Matsuoka}
  et~al.}{2019}]{mats19}
{Matsuoka} Y.,  et~al., 2019, \mn@doi [\apjl] {10.3847/2041-8213/ab0216}, \href
  {https://ui.adsabs.harvard.edu/abs/2019ApJ...872L...2M} {872, L2}

\bibitem[\protect\citeauthoryear{{Mortlock} et~al.,}{{Mortlock}
  et~al.}{2011}]{mort11}
{Mortlock} D.~J.,  et~al., 2011, \mn@doi [\nat] {10.1038/nature10159}, \href
  {http://adsabs.harvard.edu/abs/2011Natur.474..616M} {474, 616}

\bibitem[\protect\citeauthoryear{{Natarajan}, {Pacucci}, {Ferrara}, {Agarwal},
  {Ricarte}, {Zackrisson}  \& {Cappelluti}}{{Natarajan} et~al.}{2017}]{nat17}
{Natarajan} P.,  {Pacucci} F.,  {Ferrara} A.,  {Agarwal} B.,  {Ricarte} A.,
  {Zackrisson} E.,   {Cappelluti} N.,  2017, \mn@doi [\apj]
  {10.3847/1538-4357/aa6330}, \href
  {http://adsabs.harvard.edu/abs/2017ApJ...838..117N} {838, 117}

\bibitem[\protect\citeauthoryear{{Oh} \& {Haiman}}{{Oh} \&
  {Haiman}}{2002}]{oh02}
{Oh} S.~P.,  {Haiman} Z.,  2002, \mn@doi [\apj] {10.1086/339393}, \href
  {http://adsabs.harvard.edu/abs/2002ApJ...569..558O} {569, 558}

\bibitem[\protect\citeauthoryear{{Pacucci}, {Ferrara}, {Volonteri}  \&
  {Dubus}}{{Pacucci} et~al.}{2015}]{pac15}
{Pacucci} F.,  {Ferrara} A.,  {Volonteri} M.,   {Dubus} G.,  2015, \mn@doi
  [\mnras] {10.1093/mnras/stv2196}, \href
  {http://adsabs.harvard.edu/abs/2015MNRAS.454.3771P} {454, 3771}

\bibitem[\protect\citeauthoryear{{Pacucci}, {Natarajan}  \&
  {Ferrara}}{{Pacucci} et~al.}{2017}]{pac17}
{Pacucci} F.,  {Natarajan} P.,   {Ferrara} A.,  2017, \mn@doi [\apjl]
  {10.3847/2041-8213/835/2/L36}, \href
  {https://ui.adsabs.harvard.edu/abs/2017ApJ...835L..36P} {835, L36}

\bibitem[\protect\citeauthoryear{{Peebles}}{{Peebles}}{1969}]{pb69}
{Peebles} P.~J.~E.,  1969, \mn@doi [\apj] {10.1086/149876}, \href
  {https://ui.adsabs.harvard.edu/abs/1969ApJ...155..393P} {155, 393}

\bibitem[\protect\citeauthoryear{{Planck Collaboration} et~al.,}{{Planck
  Collaboration} et~al.}{2016}]{planck2}
{Planck Collaboration} et~al., 2016, \mn@doi [\aap]
  {10.1051/0004-6361/201525830}, \href
  {http://adsabs.harvard.edu/abs/2016A%26A...594A..13P} {594, A13}

\bibitem[\protect\citeauthoryear{{Regan} \& {Downes}}{{Regan} \&
  {Downes}}{2018a}]{rd18}
{Regan} J.~A.,  {Downes} T.~P.,  2018a, \mn@doi [\mnras]
  {10.1093/mnras/sty134}, \href
  {http://adsabs.harvard.edu/abs/2018MNRAS.475.4636R} {475, 4636}

\bibitem[\protect\citeauthoryear{{Regan} \& {Downes}}{{Regan} \&
  {Downes}}{2018b}]{rd18b}
{Regan} J.~A.,  {Downes} T.~P.,  2018b, \mn@doi [\mnras]
  {10.1093/mnras/sty1289}, \href
  {https://ui.adsabs.harvard.edu/abs/2018MNRAS.478.5037R} {478, 5037}

\bibitem[\protect\citeauthoryear{{Regan} \& {Haehnelt}}{{Regan} \&
  {Haehnelt}}{2009a}]{rh09a}
{Regan} J.~A.,  {Haehnelt} M.~G.,  2009a, \mn@doi [\mnras]
  {10.1111/j.1365-2966.2008.14088.x}, \href
  {http://adsabs.harvard.edu/abs/2009MNRAS.393..858R} {393, 858}

\bibitem[\protect\citeauthoryear{{Regan} \& {Haehnelt}}{{Regan} \&
  {Haehnelt}}{2009b}]{rh09b}
{Regan} J.~A.,  {Haehnelt} M.~G.,  2009b, \mn@doi [\mnras]
  {10.1111/j.1365-2966.2009.14579.x}, \href
  {http://adsabs.harvard.edu/abs/2009MNRAS.396..343R} {396, 343}

\bibitem[\protect\citeauthoryear{{Regan}, {Johansson}  \& {Haehnelt}}{{Regan}
  et~al.}{2014}]{rjh14}
{Regan} J.~A.,  {Johansson} P.~H.,   {Haehnelt} M.~G.,  2014, \mn@doi [\mnras]
  {10.1093/mnras/stu068}, \href
  {http://adsabs.harvard.edu/abs/2014MNRAS.439.1160R} {439, 1160}

\bibitem[\protect\citeauthoryear{{Regan}, {Wise}, {Woods}, {Downes}, {O'Shea}
  \& {Norman}}{{Regan} et~al.}{2020}]{ret20}
{Regan} J.~A.,  {Wise} J.~H.,  {Woods} T.~E.,  {Downes} T.~P.,  {O'Shea} B.~W.,
    {Norman} M.~L.,  2020, \mn@doi [The Open Journal of Astrophysics]
  {10.21105/astro.2008.08090}, \href
  {https://ui.adsabs.harvard.edu/abs/2020OJAp....3E..15R} {3, 15}

\bibitem[\protect\citeauthoryear{{Reinoso}, {Schleicher}, {Fellhauer},
  {Klessen}  \& {Boekholt}}{{Reinoso} et~al.}{2018}]{rein18}
{Reinoso} B.,  {Schleicher} D.~R.~G.,  {Fellhauer} M.,  {Klessen} R.~S.,
  {Boekholt} T.~C.~N.,  2018, \mn@doi [\aap] {10.1051/0004-6361/201732224},
  \href {http://adsabs.harvard.edu/abs/2018A%26A...614A..14R} {614, A14}

\bibitem[\protect\citeauthoryear{{Sakurai}, {Yoshida}, {Fujii}  \&
  {Hirano}}{{Sakurai} et~al.}{2017}]{sak17}
{Sakurai} Y.,  {Yoshida} N.,  {Fujii} M.~S.,   {Hirano} S.,  2017, \mn@doi
  [\mnras] {10.1093/mnras/stx2044}, \href
  {https://ui.adsabs.harvard.edu/abs/2017MNRAS.472.1677S} {472, 1677}

\bibitem[\protect\citeauthoryear{{Schauer}, {Whalen}, {Glover}  \&
  {Klessen}}{{Schauer} et~al.}{2015}]{anna15}
{Schauer} A.~T.~P.,  {Whalen} D.~J.,  {Glover} S.~C.~O.,   {Klessen} R.~S.,
  2015, \mn@doi [\mnras] {10.1093/mnras/stv2117}, \href
  {http://adsabs.harvard.edu/abs/2015MNRAS.454.2441S} {454, 2441}

\bibitem[\protect\citeauthoryear{{Schauer} et~al.,}{{Schauer}
  et~al.}{2017a}]{anna17}
{Schauer} A.~T.~P.,  et~al., 2017a, \mn@doi [\mnras] {10.1093/mnras/stx264},
  \href {http://adsabs.harvard.edu/abs/2017MNRAS.467.2288S} {467, 2288}

\bibitem[\protect\citeauthoryear{{Schauer}, {Regan}, {Glover}  \&
  {Klessen}}{{Schauer} et~al.}{2017b}]{srg17}
{Schauer} A.~T.~P.,  {Regan} J.,  {Glover} S.~C.~O.,   {Klessen} R.~S.,  2017b,
  \mn@doi [\mnras] {10.1093/mnras/stx1915}, \href
  {http://adsabs.harvard.edu/abs/2017MNRAS.471.4878S} {471, 4878}

\bibitem[\protect\citeauthoryear{{Shang}, {Bryan}  \& {Haiman}}{{Shang}
  et~al.}{2010}]{sbh10}
{Shang} C.,  {Bryan} G.~L.,   {Haiman} Z.,  2010, \mn@doi [\mnras]
  {10.1111/j.1365-2966.2009.15960.x}, \href
  {http://adsabs.harvard.edu/abs/2010MNRAS.402.1249S} {402, 1249}

\bibitem[\protect\citeauthoryear{{Sharda}, {Federrath}  \& {Krumholz}}{{Sharda}
  et~al.}{2020}]{sharda20}
{Sharda} P.,  {Federrath} C.,   {Krumholz} M.~R.,  2020, \mn@doi [\mnras]
  {10.1093/mnras/staa1926}, \href
  {https://ui.adsabs.harvard.edu/abs/2020MNRAS.497..336S} {497, 336}

\bibitem[\protect\citeauthoryear{{Shlosman}, {Choi}, {Begelman}  \&
  {Nagamine}}{{Shlosman} et~al.}{2016}]{shlos16}
{Shlosman} I.,  {Choi} J.-H.,  {Begelman} M.~C.,   {Nagamine} K.,  2016,
  \mn@doi [\mnras] {10.1093/mnras/stv2700}, \href
  {http://adsabs.harvard.edu/abs/2016MNRAS.456..500S} {456, 500}

\bibitem[\protect\citeauthoryear{{Smidt}, {Whalen}, {Johnson}, {Surace}  \&
  {Li}}{{Smidt} et~al.}{2018}]{smidt18}
{Smidt} J.,  {Whalen} D.~J.,  {Johnson} J.~L.,  {Surace} M.,   {Li} H.,  2018,
  \mn@doi [\apj] {10.3847/1538-4357/aad7b8}, \href
  {http://adsabs.harvard.edu/abs/2018ApJ...865..126S} {865, 126}

\bibitem[\protect\citeauthoryear{{Smith} \& {Bromm}}{{Smith} \&
  {Bromm}}{2019}]{sb19}
{Smith} A.,  {Bromm} V.,  2019, \mn@doi [Contemporary Physics]
  {10.1080/00107514.2019.1615715}, \href
  {https://ui.adsabs.harvard.edu/abs/2019ConPh..60..111S} {60, 111}

\bibitem[\protect\citeauthoryear{Smith \& Lang}{Smith \& Lang}{2019}]{ytree}
Smith B.~D.,  Lang M.,  2019, \mn@doi [Journal of Open Source Software]
  {10.21105/joss.01881}, 4, 1881

\bibitem[\protect\citeauthoryear{{Smith}, {Becerra}, {Bromm}  \&
  {Hernquist}}{{Smith} et~al.}{2017}]{aaron17}
{Smith} A.,  {Becerra} F.,  {Bromm} V.,   {Hernquist} L.,  2017, \mn@doi
  [\mnras] {10.1093/mnras/stx1993}, \href
  {http://adsabs.harvard.edu/abs/2017MNRAS.472..205S} {472, 205}

\bibitem[\protect\citeauthoryear{{Smith}, {Regan}, {Downes}, {Norman}, {O'Shea}
   \& {Wise}}{{Smith} et~al.}{2018}]{srd18}
{Smith} B.~D.,  {Regan} J.~A.,  {Downes} T.~P.,  {Norman} M.~L.,  {O'Shea}
  B.~W.,   {Wise} J.~H.,  2018, \mn@doi [\mnras] {10.1093/mnras/sty2103}, \href
  {http://adsabs.harvard.edu/abs/2018MNRAS.480.3762S} {480, 3762}

\bibitem[\protect\citeauthoryear{{Stacy}, {Bromm}  \& {Loeb}}{{Stacy}
  et~al.}{2011}]{stacy11a}
{Stacy} A.,  {Bromm} V.,   {Loeb} A.,  2011, \mn@doi [\apjl]
  {10.1088/2041-8205/730/1/L1}, \href
  {http://adsabs.harvard.edu/abs/2011ApJ...730L...1S} {730, L1}

\bibitem[\protect\citeauthoryear{{Suazo}, {Prieto}, {Escala}  \&
  {Schleicher}}{{Suazo} et~al.}{2019}]{suaz19}
{Suazo} M.,  {Prieto} J.,  {Escala} A.,   {Schleicher} D. R.~G.,  2019, \mn@doi
  [\apj] {10.3847/1538-4357/ab45eb}, \href
  {https://ui.adsabs.harvard.edu/abs/2019ApJ...885..127S} {885, 127}

\bibitem[\protect\citeauthoryear{{Surace} et~al.,}{{Surace}
  et~al.}{2018}]{sur18a}
{Surace} M.,  et~al., 2018, \mn@doi [\apjl] {10.3847/2041-8213/aaf80d}, \href
  {https://ui.adsabs.harvard.edu/abs/2018ApJ...869L..39S} {869, L39}

\bibitem[\protect\citeauthoryear{{Surace}, {Zackrisson}, {Whalen}, {Hartwig},
  {Glover}, {Woods}, {Heger}  \& {Glover}}{{Surace} et~al.}{2019}]{sur19a}
{Surace} M.,  {Zackrisson} E.,  {Whalen} D.~J.,  {Hartwig} T.,  {Glover}
  S.~C.~O.,  {Woods} T.~E.,  {Heger} A.,   {Glover} S.~C.~O.,  2019, \mn@doi
  [\mnras] {10.1093/mnras/stz1956}, \href
  {https://ui.adsabs.harvard.edu/abs/2019MNRAS.488.3995S} {488, 3995}

\bibitem[\protect\citeauthoryear{{Toomre}}{{Toomre}}{1964}]{toomre64}
{Toomre} A.,  1964, \mn@doi [\apj] {10.1086/147861}, \href
  {https://ui.adsabs.harvard.edu/abs/1964ApJ...139.1217T} {139, 1217}

\bibitem[\protect\citeauthoryear{{Toro}, {Spruce}  \& {Speares}}{{Toro}
  et~al.}{1994}]{toro94}
{Toro} E.~F.,  {Spruce} M.,   {Speares} W.,  1994, \mn@doi [Shock Waves]
  {10.1007/BF01414629}, \href
  {http://adsabs.harvard.edu/abs/1994ShWav...4...25T} {4, 25}

\bibitem[\protect\citeauthoryear{{Truelove}, {Klein}, {McKee}, {Holliman},
  {Howell}  \& {Greenough}}{{Truelove} et~al.}{1997}]{true97}
{Truelove} J.~K.,  {Klein} R.~I.,  {McKee} C.~F.,  {Holliman} II J.~H.,
  {Howell} L.~H.,   {Greenough} J.~A.,  1997, \mn@doi [\apjl] {10.1086/310975},
  \href {http://adsabs.harvard.edu/abs/1997ApJ...489L.179T} {489, L179}

\bibitem[\protect\citeauthoryear{{Tseliakhovich} \& {Hirata}}{{Tseliakhovich}
  \& {Hirata}}{2010}]{th10}
{Tseliakhovich} D.,  {Hirata} C.,  2010, \mn@doi [\prd]
  {10.1103/PhysRevD.82.083520}, \href
  {http://adsabs.harvard.edu/abs/2010PhRvD..82h3520T} {82, 083520}

\bibitem[\protect\citeauthoryear{{Turk}, {Oishi}, {Abel}  \& {Bryan}}{{Turk}
  et~al.}{2012}]{turk12}
{Turk} M.~J.,  {Oishi} J.~S.,  {Abel} T.,   {Bryan} G.~L.,  2012, \mn@doi
  [\apj] {10.1088/0004-637X/745/2/154}, \href
  {https://ui.adsabs.harvard.edu/abs/2012ApJ...745..154T} {745, 154}

\bibitem[\protect\citeauthoryear{{Umeda}, {Hosokawa}, {Omukai}  \&
  {Yoshida}}{{Umeda} et~al.}{2016}]{um16}
{Umeda} H.,  {Hosokawa} T.,  {Omukai} K.,   {Yoshida} N.,  2016, \mn@doi
  [\apjl] {10.3847/2041-8205/830/2/L34}, \href
  {http://adsabs.harvard.edu/abs/2016ApJ...830L..34U} {830, L34}

\bibitem[\protect\citeauthoryear{{Valentini}, {Gallerani}  \&
  {Ferrara}}{{Valentini} et~al.}{2021}]{vgf21}
{Valentini} M.,  {Gallerani} S.,   {Ferrara} A.,  2021, \mn@doi [\mnras]
  {10.1093/mnras/stab1992}, \href
  {https://ui.adsabs.harvard.edu/abs/2021MNRAS.507....1V} {507, 1}

\bibitem[\protect\citeauthoryear{{Vikaeus}, {Whalen}  \&
  {Zackrisson}}{{Vikaeus} et~al.}{2022}]{vik22a}
{Vikaeus} A.,  {Whalen} D.~J.,   {Zackrisson} E.,  2022, \mn@doi [\apjl]
  {10.3847/2041-8213/ac7802}, \href
  {https://ui.adsabs.harvard.edu/abs/2022ApJ...933L...8V} {933, L8}

\bibitem[\protect\citeauthoryear{{Wang} et~al.,}{{Wang} et~al.}{2021}]{wang21}
{Wang} F.,  et~al., 2021, \mn@doi [\apjl] {10.3847/2041-8213/abd8c6}, \href
  {https://ui.adsabs.harvard.edu/abs/2021ApJ...907L...1W} {907, L1}

\bibitem[\protect\citeauthoryear{{Weaver}, {Zimmerman}  \& {Woosley}}{{Weaver}
  et~al.}{1978}]{kep1}
{Weaver} T.~A.,  {Zimmerman} G.~B.,   {Woosley} S.~E.,  1978, \mn@doi [\apj]
  {10.1086/156569}, \href {http://adsabs.harvard.edu/abs/1978ApJ...225.1021W}
  {225, 1021}

\bibitem[\protect\citeauthoryear{{Whalen} \& {Fryer}}{{Whalen} \&
  {Fryer}}{2012}]{wf12}
{Whalen} D.~J.,  {Fryer} C.~L.,  2012, \mn@doi [\apjl]
  {10.1088/2041-8205/756/1/L19}, \href
  {http://adsabs.harvard.edu/abs/2012ApJ...756L..19W} {756, L19}

\bibitem[\protect\citeauthoryear{{Whalen}, {Abel}  \& {Norman}}{{Whalen}
  et~al.}{2004}]{wan04}
{Whalen} D.,  {Abel} T.,   {Norman} M.~L.,  2004, \mn@doi [\apj]
  {10.1086/421548}, \href {http://adsabs.harvard.edu/abs/2004ApJ...610...14W}
  {610, 14}

\bibitem[\protect\citeauthoryear{{Whalen}, {Mezcua}, {Meiksin}, {Hartwig}  \&
  {Latif}}{{Whalen} et~al.}{2020a}]{wet20a}
{Whalen} D.~J.,  {Mezcua} M.,  {Meiksin} A.,  {Hartwig} T.,   {Latif} M.~A.,
  2020a, \mn@doi [\apjl] {10.3847/2041-8213/ab9a30}, \href
  {https://ui.adsabs.harvard.edu/abs/2020ApJ...896L..45W} {896, L45}

\bibitem[\protect\citeauthoryear{{Whalen}, {Surace}, {Bernhardt}, {Zackrisson},
  {Pacucci}, {Ziegler}  \& {Hirschmann}}{{Whalen} et~al.}{2020b}]{wet20b}
{Whalen} D.~J.,  {Surace} M.,  {Bernhardt} C.,  {Zackrisson} E.,  {Pacucci} F.,
   {Ziegler} B.,   {Hirschmann} M.,  2020b, \mn@doi [\apjl]
  {10.3847/2041-8213/ab9d29}, \href
  {https://ui.adsabs.harvard.edu/abs/2020ApJ...897L..16W} {897, L16}

\bibitem[\protect\citeauthoryear{{Whalen}, {Mezcua}, {Patrick}, {Meiksin}  \&
  {Latif}}{{Whalen} et~al.}{2021}]{wet21a}
{Whalen} D.~J.,  {Mezcua} M.,  {Patrick} S.~J.,  {Meiksin} A.,   {Latif} M.~A.,
   2021, \mn@doi [\apjl] {10.3847/2041-8213/ac35e6}, \href
  {https://ui.adsabs.harvard.edu/abs/2021ApJ...922L..39W} {922, L39}

\bibitem[\protect\citeauthoryear{{Wise}, {Turk}  \& {Abel}}{{Wise}
  et~al.}{2008}]{wta08}
{Wise} J.~H.,  {Turk} M.~J.,   {Abel} T.,  2008, \mn@doi [\apj]
  {10.1086/588209}, \href {http://adsabs.harvard.edu/abs/2008ApJ...682..745W}
  {682, 745}

\bibitem[\protect\citeauthoryear{{Woods}, {Heger}, {Whalen}, {Haemmerl{\'e}}
  \& {Klessen}}{{Woods} et~al.}{2017}]{tyr17}
{Woods} T.~E.,  {Heger} A.,  {Whalen} D.~J.,  {Haemmerl{\'e}} L.,   {Klessen}
  R.~S.,  2017, \mn@doi [\apjl] {10.3847/2041-8213/aa7412}, \href
  {http://adsabs.harvard.edu/abs/2017ApJ...842L...6W} {842, L6}

\bibitem[\protect\citeauthoryear{{Woods} et~al.,}{{Woods}
  et~al.}{2019}]{titans}
{Woods} T.~E.,  et~al., 2019, \mn@doi [Publications of the Astronomical Society
  of Australia] {10.1017/pasa.2019.14}, \href
  {https://ui.adsabs.harvard.edu/abs/2019PASA...36...27W} {36, e027}

\bibitem[\protect\citeauthoryear{{Woods}, {Patrick}, {Whalen}  \&
  {Heger}}{{Woods} et~al.}{2021a}]{tyr22a}
{Woods} T.~E.,  {Patrick} S.,  {Whalen} D.~J.,   {Heger} A.,  2021a, arXiv
  e-prints, \href {https://ui.adsabs.harvard.edu/abs/2021arXiv211209142W} {p.
  arXiv:2112.09142}

\bibitem[\protect\citeauthoryear{{Woods}, {Patrick}, {Elford}, {Whalen}  \&
  {Heger}}{{Woods} et~al.}{2021b}]{tyr21a}
{Woods} T.~E.,  {Patrick} S.,  {Elford} J.~S.,  {Whalen} D.~J.,   {Heger} A.,
  2021b, \mn@doi [\apj] {10.3847/1538-4357/abfaf9}, \href
  {https://ui.adsabs.harvard.edu/abs/2021ApJ...915..110W} {915, 110}

\bibitem[\protect\citeauthoryear{{Woodward} \& {Colella}}{{Woodward} \&
  {Colella}}{1984}]{wc84}
{Woodward} P.,  {Colella} P.,  1984, \mn@doi [Journal of Computational Physics]
  {10.1016/0021-9991(84)90142-6}, \href
  {http://adsabs.harvard.edu/abs/1984JCoPh..54..115W} {54, 115}

\bibitem[\protect\citeauthoryear{{Woosley}, {Heger}  \& {Weaver}}{{Woosley}
  et~al.}{2002}]{kep2}
{Woosley} S.~E.,  {Heger} A.,   {Weaver} T.~A.,  2002, \mn@doi [Reviews of
  Modern Physics] {10.1103/RevModPhys.74.1015}, \href
  {http://adsabs.harvard.edu/abs/2002RvMP...74.1015W} {74, 1015}

\bibitem[\protect\citeauthoryear{{Yang} et~al.,}{{Yang} et~al.}{2020}]{yang20}
{Yang} J.,  et~al., 2020, \mn@doi [\apjl] {10.3847/2041-8213/ab9c26}, \href
  {https://ui.adsabs.harvard.edu/abs/2020ApJ...897L..14Y} {897, L14}

\bibitem[\protect\citeauthoryear{{Yue}, {Ferrara}, {Salvaterra}, {Xu}  \&
  {Chen}}{{Yue} et~al.}{2014}]{yue14}
{Yue} B.,  {Ferrara} A.,  {Salvaterra} R.,  {Xu} Y.,   {Chen} X.,  2014,
  \mn@doi [\mnras] {10.1093/mnras/stu351}, \href
  {http://adsabs.harvard.edu/abs/2014MNRAS.440.1263Y} {440, 1263}

\bibitem[\protect\citeauthoryear{{Zhu}, {Li}, {Li}, {Maji}, {Yajima},
  {Schneider}  \& {Hernquist}}{{Zhu} et~al.}{2020}]{zhu20}
{Zhu} Q.,  {Li} Y.,  {Li} Y.,  {Maji} M.,  {Yajima} H.,  {Schneider} R.,
  {Hernquist} L.,  2020, arXiv e-prints, \href
  {https://ui.adsabs.harvard.edu/abs/2020arXiv201201458Z} {p. arXiv:2012.01458}

\makeatother
\end{thebibliography}

\bsp	
\label{lastpage}
\end{document}